\documentclass[graybox]{svmult}

\usepackage{type1cm}        
\usepackage{makeidx}         
\usepackage{graphicx}        
\usepackage{multicol}        
\usepackage[bottom]{footmisc}

\usepackage{newtxtext}       %
\usepackage{newtxmath}       

\makeindex             

\usepackage{todonotes}

\begin{document}

\title*{Statistical method for analysis of interactions between chosen protein and chondroitin sulfate in an aqueous  environment}
\author{PIOTR WEBER , PIOTR BEŁDOWSKI, ADAM GADOMSKI, KRZYSZTOF DOMINO, PIOTR SIONKOWSKI, DAMIAN LEDZIŃSKI}
\institute{Piotr Weber \at Gdańsk University of Technology, G. Narutowicza 11/12, 80-233 Gdańsk \newline  \email{piotr.weber@pg.edu.pl}
\and Piotr Bełdowski \at Bydgoszcz University of Science and Technology,  Kaliskiego 7,
85-796 Bydgoszcz \newline \email{piobel000@pbs.edu.pl}
\and Adam Gadomski \at Bydgoszcz University of Science  and Technology,  Kaliskiego 7,
85-796 Bydgoszcz \email{agad@pbs.edu.pl}
\and Krzysztof Domino \at Institute of Theoretical and Applied Informatics, Polish Academy of Sciences, Bałtycka 5, Gliwice \email{kdomino@iitis.pl}
\and Piotr Sionkowski \at Institute of Theoretical and Applied Informatics, Polish Academy of Sciences, Bałtycka 5, Gliwice \email{piotr.sionkowski@gmail.com}
\and Damian Ledziński \at Bydgoszcz University of Science and Technology,  Kaliskiego 7,
85-796 Bydgoszcz \email{ damian.ledzinski@pbs.edu.pl}}
%
%
\maketitle

\section*{Abstract}
We present the statistical method to study the interaction between a chosen protein and another molecule (e.g., both being components of lubricin found in synovial fluid) in a water environment. The research is performed on the example of univariate time series of chosen features of the dynamics of mucin, which interact with chondroitin sulfate (4 and 6) in four different saline solutions. 
Our statistical approach is based on recurrence methods to analyze chosen features of molecular dynamics. Such recurrence methods are usually applied to reconstruct the evolution of a molecular system in its reduced phase space, where the most important variables in the process are taken into account. In detail, the analyzed time-series are spitted onto sub-series of records that are expected to carry meaningful information about the system of molecules. Elements of sub-series are splinted by the constant delay-time lag (that is the parameter determined by statistical testing in our case), and the length of sub-series is the embedded dimension parameter (using the Cao method). We use the recurrent plots approach combined with the Shannon entropy approach to analyze the robustness of the sub-series determination. We hypothesize that the robustness of the sub-series determines some specifics of the dynamics of the system of molecules. We analyze rather highly noised features to demonstrate that such features lead to recurrence plots that graphically look similar. From the recurrence plots, the Shannon entropy has been computed. We have, however, demonstrated that the Shannon entropy value is highly dependent on the delay time value for analyzed features. Hence, elaboration of a more precise method of the recurrence plot analysis is required. For this reason, we suggest the random walk method that can be applied to analyze the recurrence plots automatically.

\section*{Keywords}
embedded dimension; recurrence plots; statistical testing; Shannon entropy;
synovial fluid interactions; molecular docking numerical experiment.

\section{Introduction}
\label{sec:1}

The motivation of the research comes from the need to elaborate statistical methods dedicated to analyzing components of synovial fluid present at articular cartilage. 
This chapter concentrates on the method dedicated to analyzing data from chemical polymeric system and synovial fluid (SF) dynamics.
Generally, complex systems, such as those mentioned above,
are characterized by non-linear dynamics and diffusion regime~\cite{ben2000diffusion} that leads to a random walk with long-range memory, governed by the fractional master equation~\cite{jumarie2001fractional}. Moreover, such systems are often stochastic in that 
they are governed partly by the deterministic process and partially by its stochastic counterpart. Finally, it is essential to mention that anomalous (sub)diffusion may be expected to manifest for cellular membranes and lipid bilayer \cite{kruszewska2020interactions}, but its investigation is not straightforward  \cite{saxton2012wanted}.
We apply and further evaluate the state of art methods applicable to analyze these types of stochastic data, i.e., determination of the auto-correlation lag (delay time $\tau$), determination of the embedded dimension $d$ (by the Cao method~\cite{Cao}), application the Shannon entropy approach~\cite{MarwanRomanoThielKurths2007}, and the perforation of recurrence plots~\cite{eckmann1995recurrence}. Then we can point out similar and different dynamics of various polymers versions immersed in various salts solutions.
Our novel approach enriches the methods mentioned above by typical data engineering approaches, such as statistical testing. 
From the informative point of view, the univariate time series of data from molecular dynamics can be spitted onto many $d$ long sub-series of records separated by the $\tau$-long lag. Each sub-series is expected to carry most of the meaningful information from the original series. The comparison of this sub-series, and the analysis of the robustness of their formation, can be performed by means of the recurrence plots approach. We hypothesize here that this robustness is tied to the dynamic of the molecular system. Hence the analysis of highly noised features, such as total Van der Waals energy, should result in similar recurrence plots (reflecting dynamics of water-based noise). Further, some slight differences in recurrence plots may point out the direction for extracting the true information about the immersed molecules.

To sample data for analysis, we select an example of the study of the interactions of mucin with glycosaminoglycans by molecular docking followed by molecular dynamics simulations. Molecular docking is the numerical simulation method returning the total bond energy and the orientation of one molecule in relation to another so that they form a stable complex~\cite{PinziRastelli}. Molecular docking is based on the fundamental principle of the so-called molecular recognition of molecules, i.e., a fundamental signature of a complex system of synergistic property ~\cite{Haken}, in order to evoke a specific biological response. The receptor and ligand (in our work, these are mucin-glycosaminoglycans respectively) subjected to the calculation process take the optimal orientation for both molecules; it resembles the so-called key-and-lock principle characteristic of the complex systems. As a result, the system's free energy as a whole is minimized. The whole docking process does not only consist of searching for the optimized confirmation but also of evaluating the obtained complex based on its free energy; thus, a crucial thermodynamic quantity ~\cite{PinziRastelli}. This study investigates the interactions of mucin - glycosaminoglycans by molecular docking to obtain the total bond energy, binding contacts, and the optimized conformation and orientation of this very complex system, which is immersed in one chosen aqueous salt solution.

To place our research in biology and medicine, observe that articular cartilage at its extracellular level is composed of cartilage cells — chondrocytes submerged in the extracellular matrix (ECM) ~\cite{Klein2013,Klein2021}. One of the components of the matrix are proteoglycans, i.e., complexes of high molecular weight proteins with glycosaminoglycans - chondroitin 4-sulfate (CS-4), chondroitin 6-sulfate (CS-6), and keratan sulfate (KS) 
Glycosaminoglycans (GAG) with a protein core create spatial structures (resembling scaffoldings), which are capable of binding large amounts of water~\cite{Loret_Simoes}. 
Lubricin is present in the synovial fluid and on the joint cartilage surface. It is a proteoglycan that is formed in chondrocytes. Its central domain is mucin, which is subject to O-linked glycosylation~\cite{Dube4819} that can reach between $150-250$ nm. Mucins are a family of high molecular weight glycoprotein polymers, heavily glycosylated, expressed both as cell membrane-tethered molecules and as a major component of the mucus gel~\cite{BROWN2013200}. Mucins play an important role in the regulatory mechanisms of many species~\cite{BROWN2013200}. The multiple and diverse physiological roles of mucins have long been of interest. However, some of their biophysical and biological properties have been investigated in more detail only more recently~\cite{BROWN2013200}. 
One of the vital meaning properties attributed to mucins is lubrication~\cite{Kasdorf_Weber}.
Lubrication of articular cartilage is due to the synovial fluid that contains up to 70\% of water, proteins (albumin and gamma-globulin), phospholipids, and biopolymers. Such lubrication is a very efficient process that is achieved as a result of interactions between SF components; complexes of two or more SF components show up a lower coefficient of friction than these components individually~\cite{seror2015supramolecular,Haken}.
The lubrication synergy of DPPC bilayers covered by hyaluronan was investigated in~\cite{raj2017lubrication}, as these bilayers give rise to remarkably low friction coefficients. (For further discussion on this low friction phenomenon, see~\cite{C1SM06335A}, ~\cite{dedinaite2017synergies}.) The probable cause of this important phenomenon is the specific form of biomolecular  aggregation~\cite{oates2006rheopexy}. In~\cite{gadomski2008directed} the hypothetical model of the friction–lubrication mechanism in articular cartilage, based on an anomalous (first-order) chemical reaction involving lipids in association with hyaluronan (and mediated by water) was presented \cite{gadomski2013}.
The complex and not fully explained mechanism of cartilage lubrication provides proper motivation for analyzing the dynamics of synovial fluid and its components. 





We divide this chapter into sections, where we present the methodology for analyzing the interaction of mucin with one of the glycosaminoglycan (CS-4 or CS-6) in a selected aqueous solution of one salt.
In Section~\ref{sec:2} we discuss the methodology of data processing using in recurrence method (delay time, embedded dimension), which we want to use here. In Section~\ref{sec:3} we discuss recurrence plots  and entropy calculations. In Section~\ref{sec::results} we perform the comparative analysis of the results and in Section~\ref{sec::conclusions} we supply conclusions.

\section{Data processing - delay time and embedded dimension}
\label{sec:2}

\begin{figure}
\centering
    \includegraphics[width=1.0\textwidth]{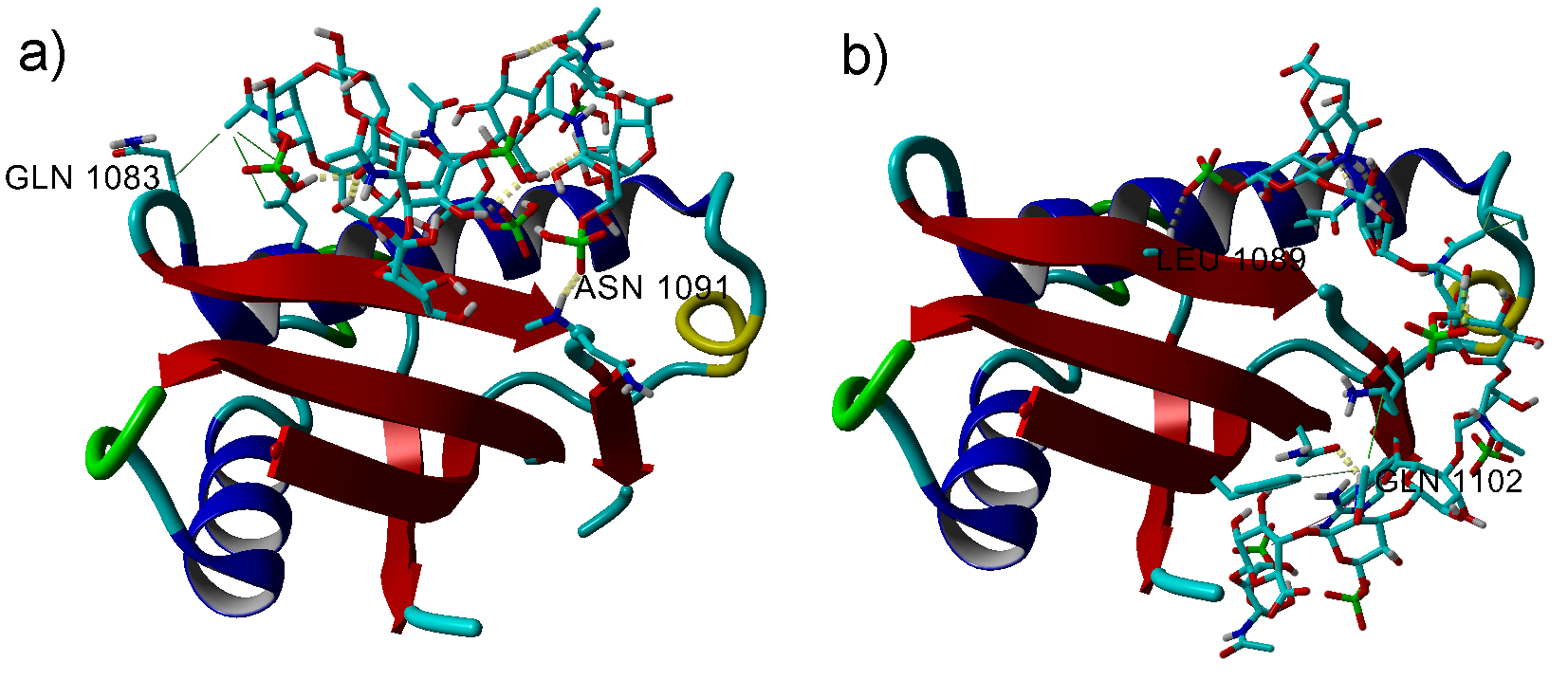}
    \caption{ Snapshot from docking experiment for best conformations of chondroitin sulfate and mucin complexes: a) CS-4, b) CS-6. }\label{fig::snapshot}
\end{figure}

At first, we discuss how time series for analysis are obtained. Then, we analyze the dynamics of the vital component of the synovial fluid, the mucin. Finally, we perform numerical experiments by the YASARA Structure Software (Vienna, Austria)~\cite{krieger2015new} program. We docked complexes using the VINA method ~\cite{Trott_Olson} with default parameters and point charges initially assigned to the AMBER14 force field \cite{AMBER_14}.
In detail, we analyze mucin connected either with CS-4 or CS-6 see Fig.~\ref{fig::snapshot}. In both cases, mucin is immersed in a water solution of various salts (CaCl$_2$, KCl, MgCl$_2$ or NaCl) to reflect better particular compositions of the synovial fluid. 

Simulations were followed with data analysis of the system of interest. For the detailed analysis of the whole system, we choose - total Van der Waals energy, as this feature is highly disturbed by the water-based noise. Hence this feature is a good example to test the robustness of the proceeding method on the noise. We present the step-by-step data proceeding scheme enriched by plots for this feature. 
We analyze $K = 14$ independent molecular simulations for each experimental setting to gather data for statistical analysis. 

\begin{figure}
\centering
    \includegraphics[width=0.45\textwidth]{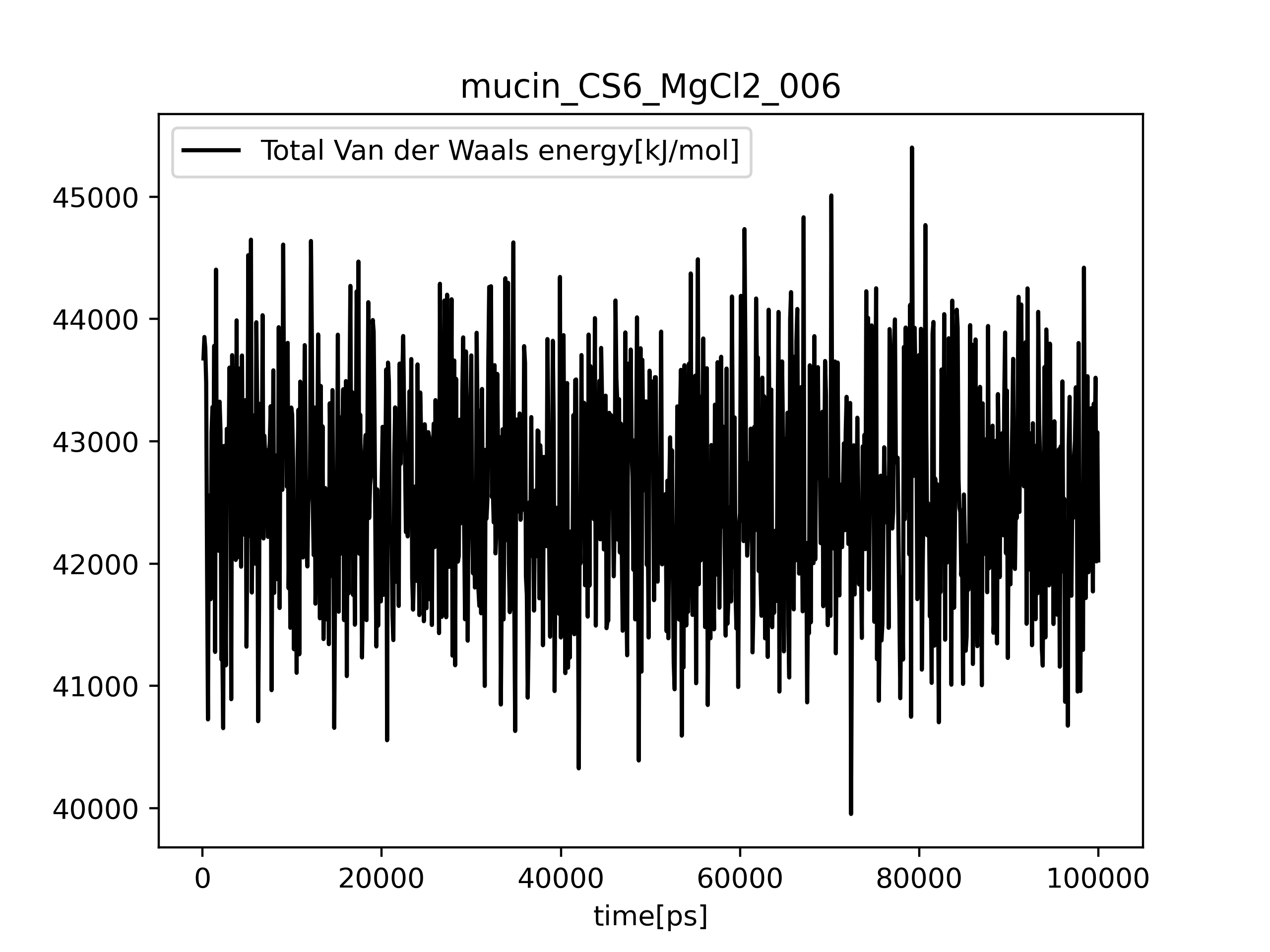}
    \includegraphics[width=0.45\textwidth]{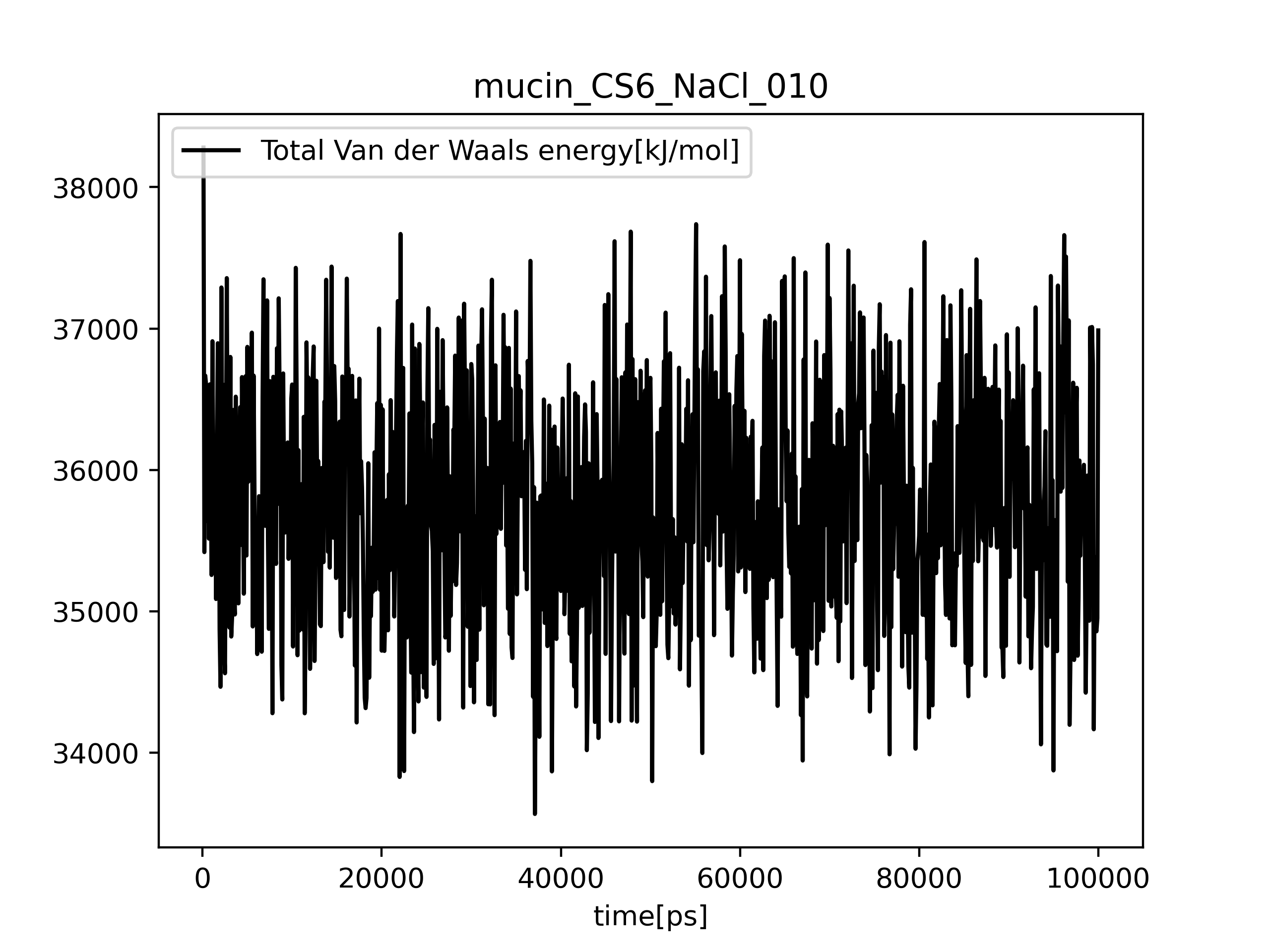}
    \caption{Pattern of analysed feature (total Van der Waals energy) for chosen setting and simulation number, observe possible cyclic component.  }\label{fig::energy}
\end{figure}

As mentioned in the introduction, we expect the analyzed signal to be stochastic, i.e., governed by the deterministic and random processes. Some data are presented in Fig.~\ref{fig::energy}. In addition, some cyclic components might be observed, perhaps because of some oscillations around the balance state of the molecular complex.

Using the records of the series, we aim to obtain a trajectory in the phase space of the system generating the signal. We apply here the recurrent plot method, which is applicable for rather short and possibly non-stationary data \cite{PhysRevA.34.4971}. This observation leads to the time delay and embedded dimension analysis.

In more details, for each molecular simulation we analyse $N$-long univariate time series
\begin{equation}
    \vec{x} = (x_1, x_2, \ldots, x_N),
\end{equation}
as these time series are dense in time, subsequent elements may carry similar information. In such case, to reduce number of data records, we can limit our-self to the sub-series that is meaningful from the informative point of view. We aim to construct following vectors:
\begin{equation}
\vec{y}_{i}^{(d, \tau)} = \left(x_{i}, x_{i+\tau}, x_{i+2 \tau},\ldots,
x_{i+d \tau} \right),
\label{vector}
\end{equation}
parameter $\tau$ is called delay time (if $\tau > 1$ some cyclic component would be filtered out by reducing the number of data by the factor of $\tau$), and parameter $d$ is called parameter embedded dimension of the system. (The estimated values  of these will be called $T$ and $d_0$.) The time evolution of this vector is realize by transformation $\vec{y}_{i}^{(d, \tau)} \rightarrow \vec{y}_{i+1}^{(d, \tau)}$, these trajectories are analyzed in our work.

\subsection{Determination of delay time}\label{sec::tau}

To estimate the delay time $\tau$ we use the method proposed in \cite{peppoloni2017characterization}\cite{goswami2019brief}\cite{subha2010eeg}, searching for such subsequent sub-series that have zero or almost zero correlation function (i.e. no/smallest mutual information). The lag between these series is expected to split elements in $\vec{y}_{i}^{(d, \tau)}$.

The simplest approach is the \emph{first zero crossing} of auto-correlation \cite{peppoloni2017characterization} \cite{goswami2019brief}\cite{subha2010eeg} adequate for rather deterministic (low noised) signals.

\par\noindent{\bf Input:}
\begin{itemize}
\item $\vec{x} \in \mathbb{R}^N$ 
\end{itemize}
\par\noindent{\bf Processing:} 
\begin{enumerate}
    \item define $\vec{x}^{(1)} = x_1, x_2, \ldots, x_{N-\tau}$ and $\vec{x}^{(2)} = x_{1+\tau}, x_{2+\tau}, \ldots, x_{N}$
    \item compute auto-correlation function:
    \begin{equation}
    a(\tau)=\text{cor}(\vec{x}^{(1)}, \vec{x}^{(2)})
    \label{autokorelacja}
    \end{equation}
    where cor is the particular measure of correlation, i.e. Pearson (for Gaussian distributed data), Spearman, Kendall, etc. We select the Spearman correlation as we can not make an assumption about the Gaussian distribution of data.
     \item determine $\tau_{-}$ such that $a(\tau_{-}) < 0$ and $\forall_{\tau \in \{0,1, \ldots,  \tau_{-} - 1\}} \ a(\tau) > 0$
    \item return \begin{equation}\tau = \text{arg min}_{\tau \in \{0,1, \ldots,  \tau_{-} - 1\}} a(\tau).
    \end{equation}
\end{enumerate}
\par\noindent{\bf Output:} $\tau$

However, this method may not be adequate if data are noisy (we may have jumps between positive and high but negative auto-correlations). For such noisy data, we propose the \emph{first not-significant auto-correlation} approach. For this reason, we replace points $3$ and $4$ of the algorithm mentioned above by statistical testing.  

\par\noindent{\bf Input:}
\begin{itemize}
\item $\vec{x} \in \mathbb{R}^N$ 
\end{itemize}
\par\noindent{\bf Processing:} 
\begin{enumerate}
    \item define $\vec{x}^{(1)} = x_1, x_2, \ldots, x_{N-\tau}$ and $\vec{x}^{(2)} = x_{1+\tau}, x_{2+\tau}, \ldots, x_{N}$
    \item compute auto-correlation function:
    \begin{equation}
    a(\tau)=\text{cor}(\vec{x}^{(1)}, \vec{x}^{(2)})
    \label{autokorelacja}
    \end{equation}
     \item perform the hypothesis testing:
     \begin{itemize}
        \item $H_0$ -  $a(\tau) = 0$,
        \item $H_1$ - $a(\tau) \neq 0$.
    \end{itemize}
    \item return the lowest $\tau$ for which $H_0$ can not be rejected at $p$ significance level, i.e. $\text{Prob}(a(\tau) = 0) >  p$. 
\end{enumerate}
\par\noindent{\bf Output:} $\tau$

\begin{figure}
\centering
    \includegraphics[width=0.49\textwidth]{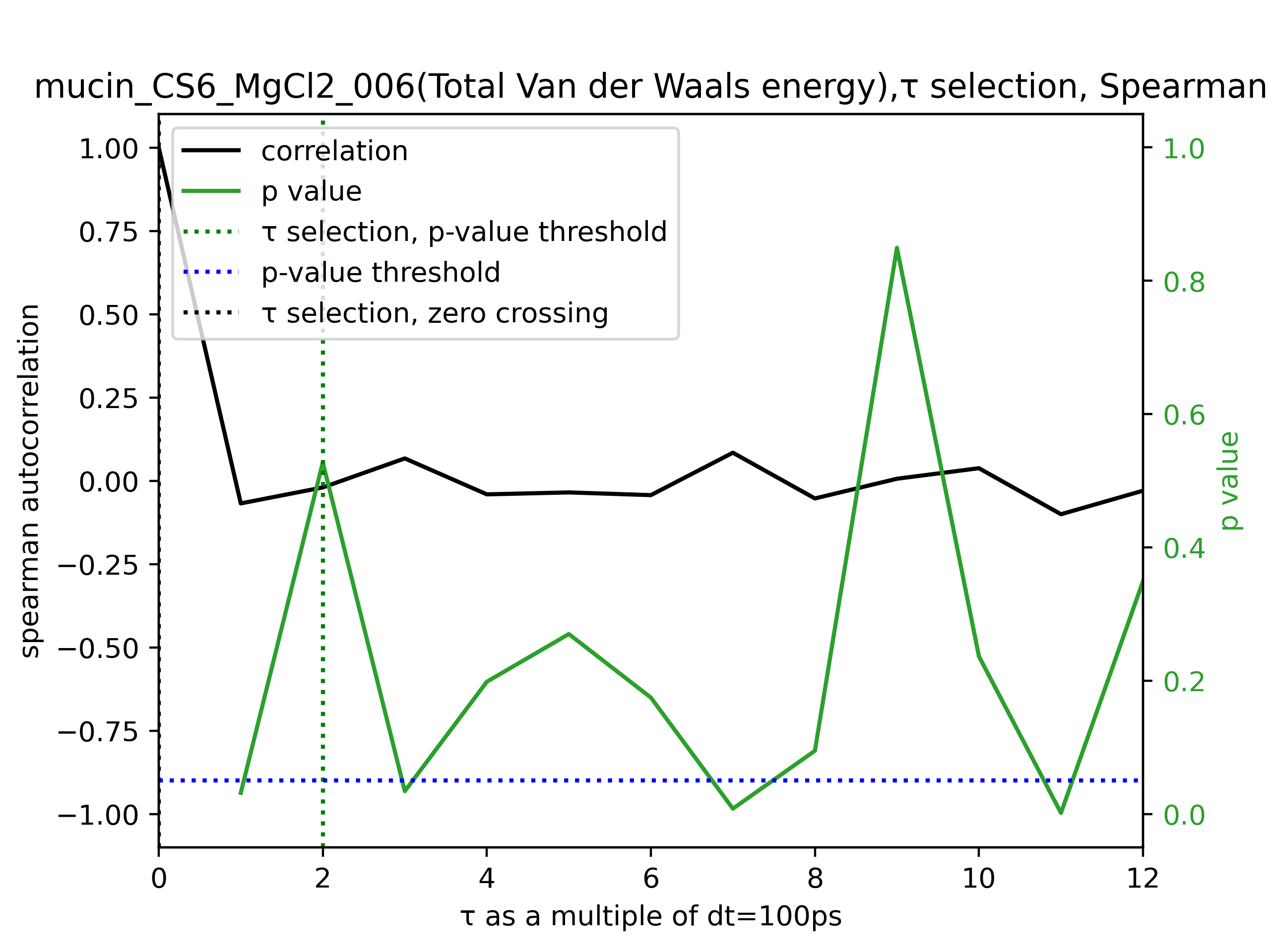}
    \includegraphics[width=0.49\textwidth]{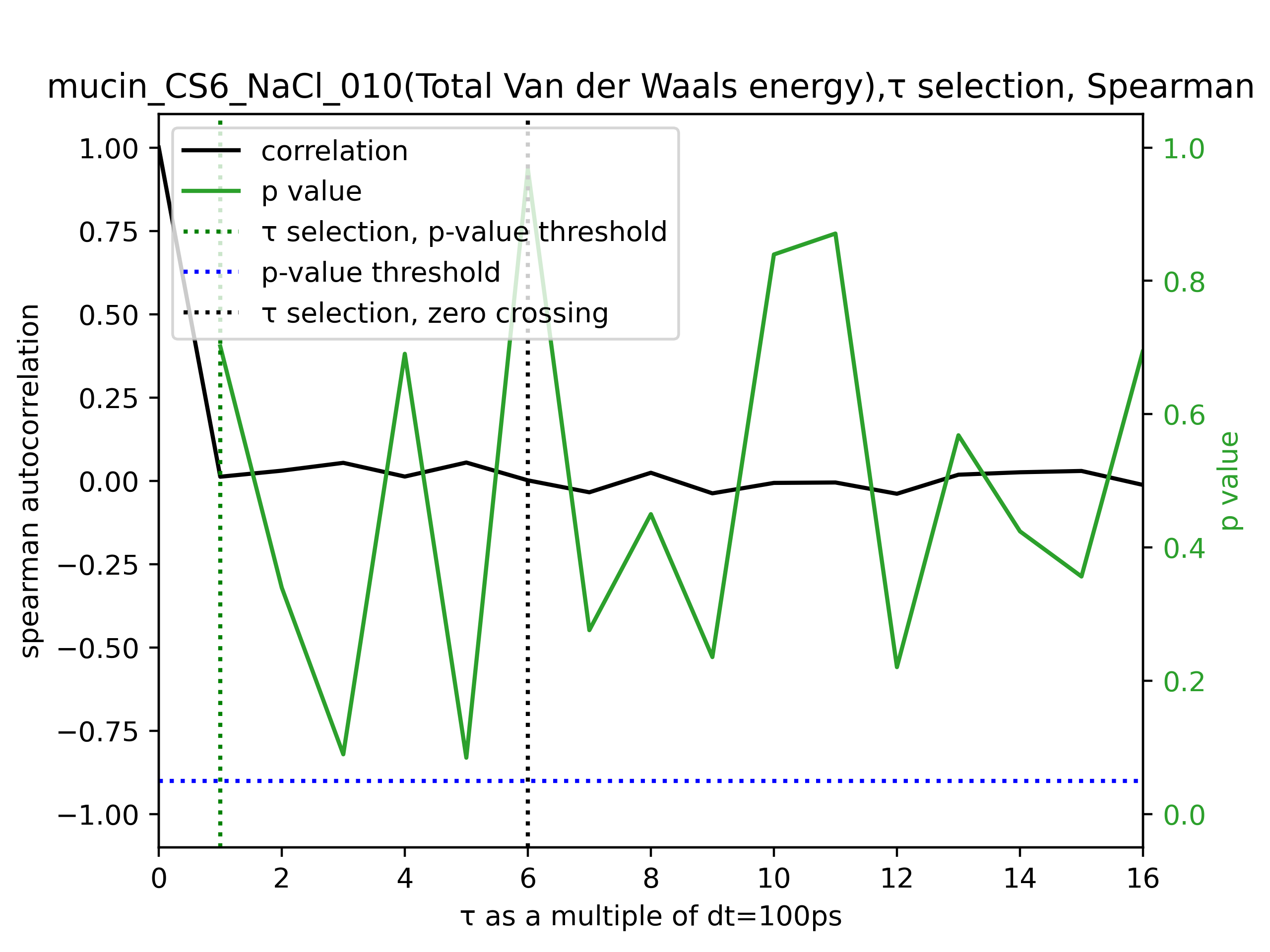}
    \caption{$\tau$ Determination of the delay-time by both \emph{first zero crossing} auto-correlation and \emph{first non-significant auto-correlation} method. Total Van der Waals energy of mucin with CS-6 was chosen in both cases. Observe that in left panel the \emph{first zero crossing} returns $\tau = 0$ what is not the result one should expect. }\label{fig::get_tu}
\end{figure}

The comparison of two above-mentioned methods are presented in Fig.~\ref{fig::get_tu} for particular experimental setting and particular experimental realisation concerning total Van der Waals energy data. We observed that sometimes (left panel) \emph{first zero crossing} method returns $\tau = 0$ since the auto-correlation for $\tau = 1$ is significantly negative. Because of this drawbacks of \emph{first zero crossing} method, we have selected \emph{first not-significant auto-correlation}  method for further investigation. This method appears to be more stable, for most investigated data of total Van der Waals energy, the method returns $\tau = 1$ or (less frequently) $\tau = 2$.

\subsection{Determination of embedded dimension}

The method uses vectors $\mathbf{y}_i^{(d, T)}$ defined by Eq.~\eqref{vector}, with already estimated $T$, having this in mind hereafter we will drop $T$ and write $\mathbf{y}_i^{(d)}$. These vectors (numbered by $i$) uniquely mark points in $d_0$-dimensional space. 
If $d > d_0$, some of $\mathbf{y}_i^{(d)}$ elements (points) would be projections from a higher dimension phase space onto actual $d_0$-dimensional space, the false neighbours. We aim to determine from data space dimensionality-$d_0$ (embedded dimension) and reduce data size to have no false neighbors.

To determine the embedded dimension, we can follow the approach introduced by Cao in ~\cite{Cao}. Using $\vec{y}_{i}^{(d)}$ (the $i$ - th reconstructed vector in $d$ dimensional phase space) and $\vec{y}_{i}^{(d+1)}$ 
one defines:
\begin{equation}
a_{(i,d)}=
\frac{\|\vec{y}_{i}^{(d+1)} - (\vec{y}_{1}^{(d+1)}, \ldots, \vec{y}_{M}^{(d+1)}) \|_{\min, \max}}{\|\vec{y}_{i}^{(d)} - (\vec{y}_{1}^{(d)}, \ldots, \vec{y}_{M}^{(d)}) \|_{\min, \max}},  \quad   i=1,2,…,M,
\label{wsp_a_i_d}
\end{equation}
where $M$ is the number of vectors that can be constructed from the signal for a given $T$ (the estimate of $\tau$) and $d+1$ dimension. 

The distance is given by:
\begin{equation}
\|\vec{y}_{i}^{(d)} - (\vec{y}_{1}^{(d)}, \ldots, \vec{y}_{M}^{(d)}) \|_{\min, \max} = 
\min_{j \in \{1:M\}: \vec{y}_j \neq \vec{y}_i} \left( \max_{1 \leq k \leq d} abs \left( {y_i}_k-{y_j}_k \right) \right). 
\label{10}
\end{equation}

Next one has to calculate $\bar{a}_{(d)}$ - the mean value of $a_{(i,d)}$ over $i$,
and investigate its increment from $d$ to $d+1$ in the ratio form:
\begin{equation}
R(d)=\frac{\bar{a}_{(d+1)}}{\bar{a}_{(d)}}.\label{12}
\end{equation}
This value grows with $d \in \{1,2, \ldots, d_0 \}$ and is roughly one for $d \in \{d_0+1, d_0+2, \ldots  \}$, hence $d_0$ can be treated
as the estimator of embedded dimension.

The computation of $R(d)$ is summarised in the following algorithm:
\par\noindent{\bf Input:}
\begin{itemize}
\item $\vec{x}$ - signal
\item $T$ - already estimated delay time
\item $d_{\max}$ - the maximal embedded dimension.
\end{itemize}
\par\noindent{\bf Processing:} 
\begin{enumerate}
    \item define
    $\vec{y}_{i}^{(d)} = (x_{i},x_{i+T},x_{i+2 T},…,x_{i+d T})$, set $M$ to be the number of elements of $\vec{y}_{i}^{(d+1)}$;
    \item for each $i \in \{1, \ldots, M\}$ we return such $\| {\vec{y}^{(d)}_i} - {\vec{y}^{(d)}_j} \|_{\max}$ that is minimal over $j \in \{ 1:i-1,i+1:M \}$  (but non-zero) 
    \item repeat $(3)$ for $\vec{y}^{(d+1)}$, return $a_{(i,d)}$ and $\bar{a}_{(d)}$;
    \item from $\bar{a}_{(d)}$ and $\bar{a}_{(d+1)}$ compute $R(d)$; 
    \item repeat procedure in $(1) - (5)$ for $d \in \{1,2, \ldots,  d_{\max} \}$;
    \item return vector $R(d)$ for $d \in \{1,2, \ldots,  d_{\max} \}$.
\end{enumerate}
\par\noindent{\bf Output:}
\begin{itemize}
\item vector $R(d)$.
\end{itemize}

\begin{figure}
\centering
    \includegraphics[width=0.49\textwidth]{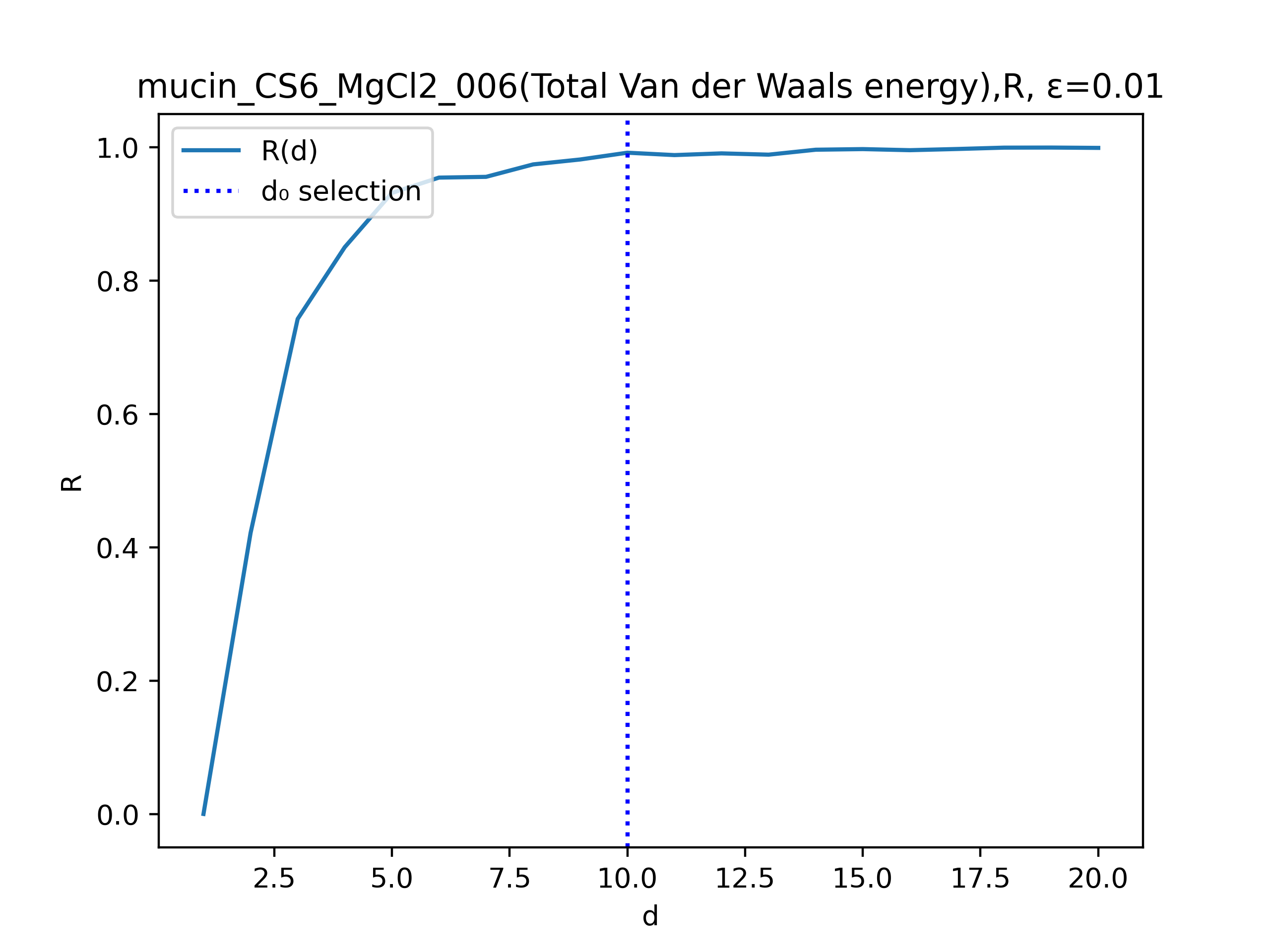}
    \includegraphics[width=0.49\textwidth]{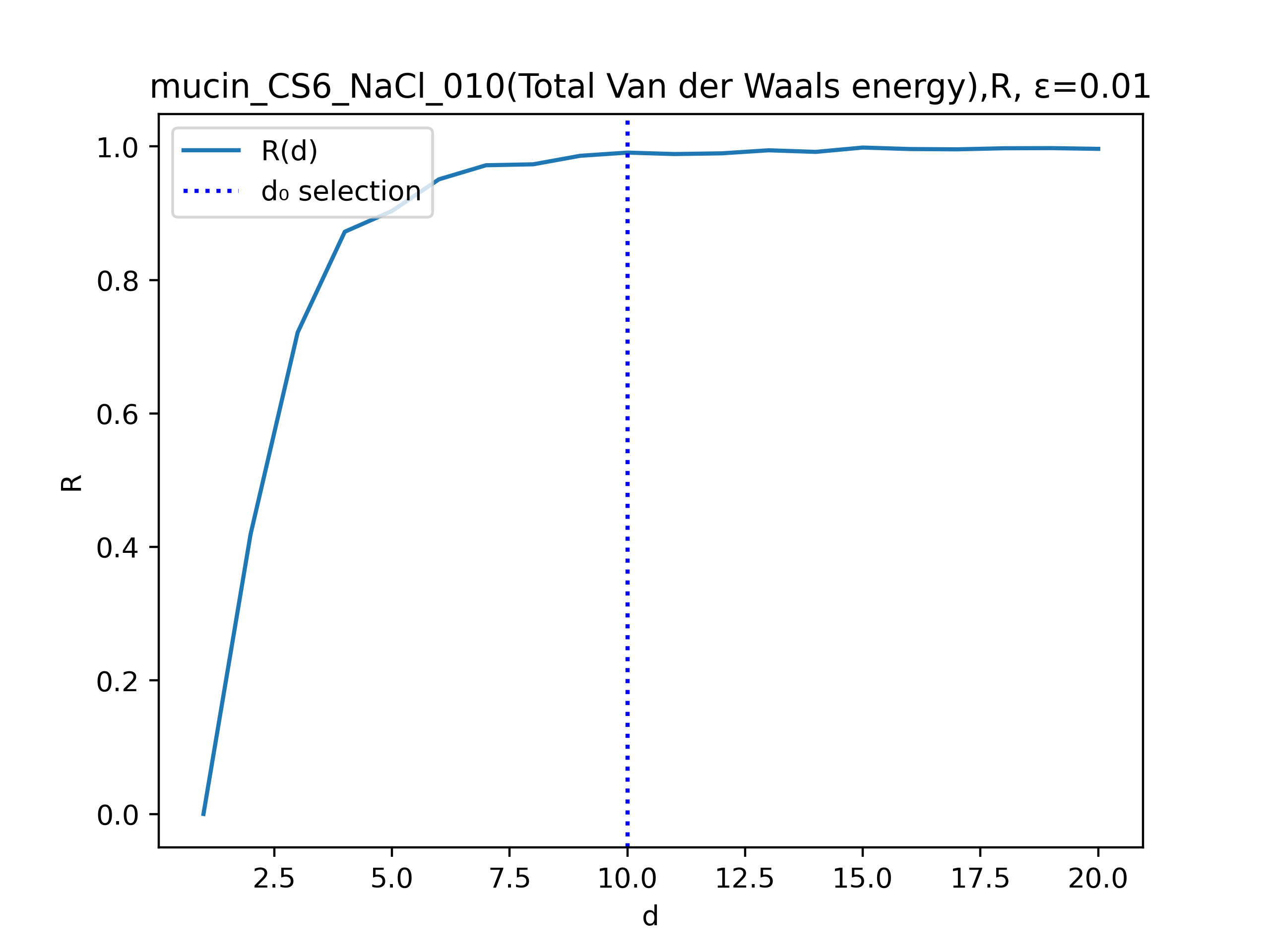}
    \caption{ Determination of embedded dimension $d_0$ via the plot of $R$, one can observe that the cutoff according to Eq.~\eqref{eq::R_cut} with $\varepsilon = 0.01$ gives expected results. Total Van der Waals energy of mucin with CS-6 was chosen in both cases. }\label{fig::get_R}
\end{figure}

Given the $R(d)$, the embedded dimension $d_0$ is the minimal value such that $R(d)$ is roughly $1$ for all $d > d_0$. In Fig~\ref{fig::get_R} we present some plots of $R(d)$ for chosen experiment settings (concerning total Van der Waals energy of mucin with CS-6 was chosen in both cases) and chosen realization. Here $d_0$ is determined by the following cutoff procedure.
To omit such subjective methodology of $d_0$ determination, we propose the automatic one that uses the threshold value $\varepsilon$ parameter. As $d_0$ we return smallest $d$ fulfilling
\begin{equation}\label{eq::R_cut}
R(d) \geq 1-\varepsilon.
\end{equation}

We present the above procedure results for total Van der Waals of mucin, CS-4, and water one chosen salt solution in the Table below.

\begin{center}
\begin{tabular}{lcccc}
Realisation & CaCl$_{2}$ & KCl  & MgCl$_2$ &  NaCl \\
\hline
\hline
001 & d=10  & d=11 &  d=10   & d=12   \\
002 & d=11  & d=12 &  d=12   & d=10   \\
003 & d=10  & d=10 &  d=9    &  d=9   \\
004 & d=11  & d=10 &  d=11   &  d=11   \\
005 & d=12  & d=11 &  d=10   &  d=12   \\
006 & d=11  & d=9  &  d=10   &  d=11  \\
007 & d=12  & d=9  &  d=10   &  d=10  \\
008 & d=10  & d=11 ($T$=2) &  d=11   &  d=12   \\
009 & d=10  & d=12 &  d=12   &  d=12   \\
010 & d=11  & d=11 &  d=13   &  d=10   \\
011 & d=11  & d=10 &  d=11   &  d=10   \\
012 & d=8   & d=11 &  d=12   &  d= 8  \\
013 & d=10  & d=12 &  d=10   &  d=10  \\
014 & d=10  & d=13 &  d=10   &  d=9 \\
\hline	
\end{tabular}
\end{center}

We present the above procedure results for total Van der Waals of mucin, CS-6, and water one chosen salt solution in the Table below.

\begin{center}
\begin{tabular}{lcccc}
Realisation & CaCl$_{2}$ & KCl  & MgCl$_2$ &  NaCl \\
\hline
\hline
001 & d=11  & d=11 ($T$=2)  & d=10  & d=9 ($T$=2)  \\
002 & d=11  & d=11             & d=11  & d=9  \\
003 & d=12  & d=12             & d=9   & d=11  \\
004 & d=11  & d=12             & d=12  & d=10 ($T$=2)  \\
005 & d=10  & d=11             & d=12  & d=11  \\
006 & d=12  & d=10             & d=10 ($T$=2)  & d=10  \\
007 & d= 9  & d=11             & d=11  & d=10  \\
008 & d=11  & d=11             & d=11  & d=9  \\
009 & d=11  & d=12             & d=11  & d=11 ($T$=2)  \\
010 & d=11  & d=11             & d=12  & d=10  \\
011 & d=11  & d=10             & d=11  & d=10  \\
012 & d=10  & d=12             & d=11  & d=12 ($T$=2)  \\
013 & d=12  & d=10             & d=9   & d=10  \\
\hline	
\end{tabular}
\end{center}
We can see that the embedded dimension is not the same for all pairs: mucin-CS-4 or mucin-CS-6, each in a single salt solution. However, this is a large reduction in the degrees of freedom of a system whose number of atoms is made up of several hundred atoms. Let us also note that for several cases, there is $T=2$; if not noted, we had $T = 1$. 

\section{Recurrence plots}
\label{sec:3}

To analyze data characteristics, we use the recurrence plots and Shannon entropy. To compute these, we use the distance matrix between the points in the reconstructed phase space~\cite{eckmann1995recurrence}. Suppose $d_0$ and $T$ have already been estimated. Then, let $\mathbf{y}^{(d_0)}_k$ and $\mathbf{y}^{(d_0)}_{k'}$ be two-point in the phase space, see Eq.~\eqref{vector}. In theory, they should be the same for the completely cyclic signal. To measure the random factor and non-cyclic trends, we use the euclidean distance:
\begin{equation}
    \delta_{k,k'} = \| \mathbf{y}_k^{(d_0)} - \mathbf{y}_{k'}^{(d_0)} \|^2
    \label{eq::delta_norm}
\end{equation}
Then we use the parameter $\lambda$ such that if $\delta_{k,k'} \leq \lambda$ the signals are regraded as similar, otherwise they are regarded as different. This can be represented graphically by matrix of zeros and ones with entries:
\begin{equation}
    m_{k, k'}(\lambda) = \begin{cases} 0 &\text{ if } \delta_{k,k'} \leq \lambda \\ 1 &\text{ elsewhere } \end{cases}
\label{eq::m_matrix}
\end{equation}

The parameter $\lambda$ is a key part of creating a recurrence plot, but there is no mathematical derivation of the particular $\lambda$ value.
There are only some constraints in the literature (that we are willing to follow). Typically ~\cite{Goswami} fraction of zeros in Eq.~\eqref{eq::m_matrix}, called also the recurrence rate:
\begin{equation}
RR(\lambda)=\frac{1}{M^{2}}\sum_{k,k'} 
\left[1 -  m_{k, k'}(\lambda) \right]
\end{equation}
should fit into the range $5\% - 10\%$.

To determine particular $\lambda$ value, we propose to maximise the entropic measure introduced in \cite{MarwanRomanoThielKurths2007}. In the matrix in Eq.~\eqref{eq::m_matrix} we analyse lines  parallel to the diagonal, searching for vectors consisting of the sequence of zeros. Then we calculate histogram of lengths of such vectors and normalise them to $1$, normalised histograms are $p_{i}(\lambda)$, such that
\begin{equation}\sum_{i=1}^M p_i(\lambda) = 1
\end{equation}
where $i=1,2,…,M$.  
The Shannon entropy is
\cite{MarwanRomanoThielKurths2007}:
\begin{equation}
S(\lambda)=-\sum_{i=1}^{M}p(\lambda)_{i}\ln p_{i}(\lambda)  
\label{eq::entropy}
\end{equation}   
From a mathematical point of view, 
the entropy value tends to zero when the probability distribution tends to the distribution where one diagonal length is most probable. On the other hand, the entropy value tends to its maximum value as the diagonal length distribution tends to a homogeneous distribution.

The algorithm for $\lambda$ determination can be pointed out as follow:
\par\noindent{\bf Input:}
\begin{itemize}
\item $\vec{x}$ - signal
\item $T$ - already estimated delay time
\item $d_{0}$ - already estimated embedded dimension.
\end{itemize}
\par\noindent{\bf Processing:} 
\begin{enumerate}
    \item compute $\vec{y}^{(d_0)}_k$ using Eq.~\eqref{vector} for $k \in 1, \ldots , M$,
    \item compute $\delta_{k,k'}$ using Eq.~\eqref{eq::delta_norm},
    \item determine $\lambda$ such that:
    \begin{itemize}
    \item $0.05 \leq RR(\lambda) \leq 0.1$
    \item maximise $S(\lambda)$
    \end{itemize}
\end{enumerate}
\par
\begin{figure}
    \centering    \includegraphics[width=0.49\textwidth]{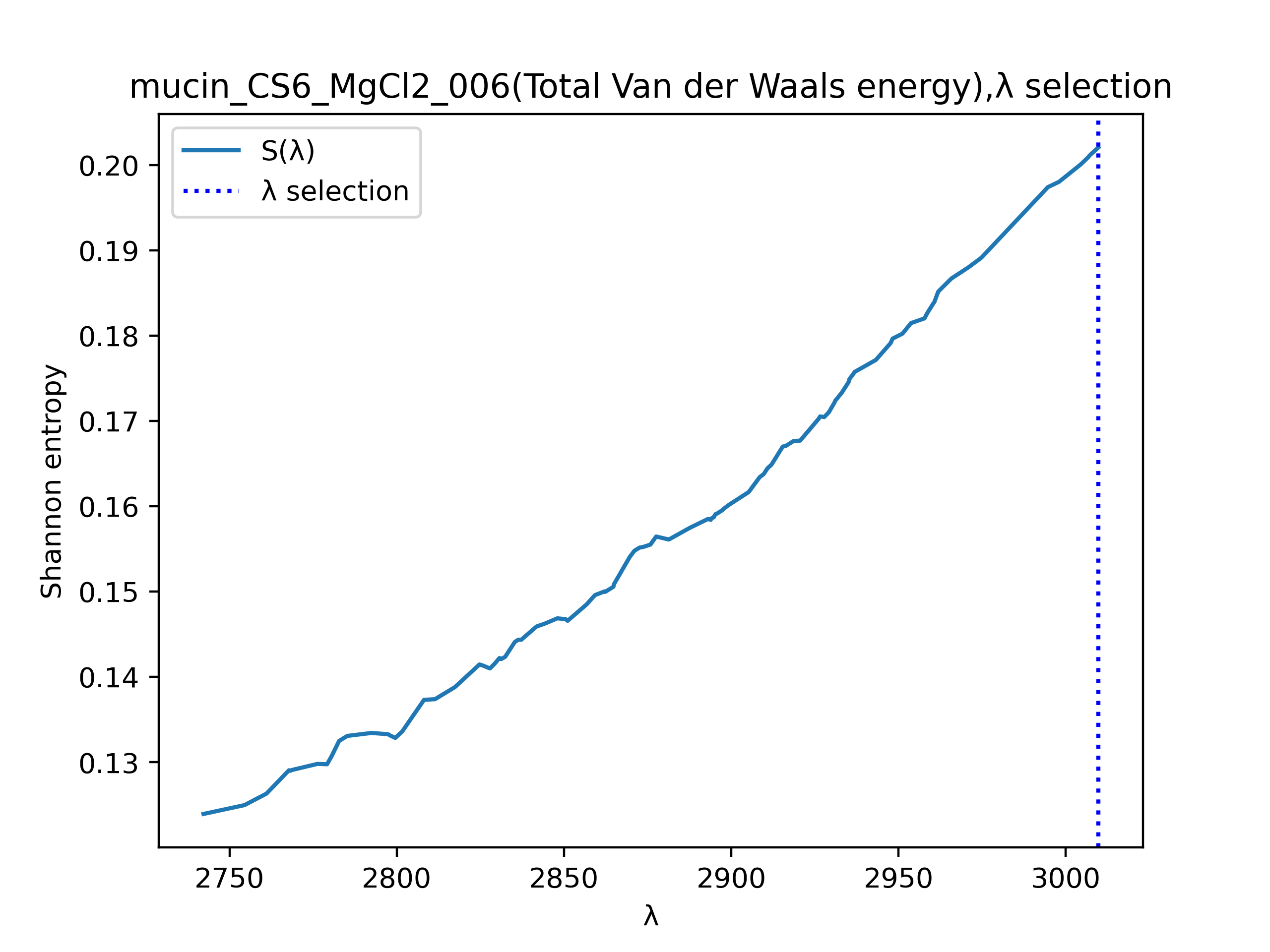}
    \includegraphics[width=0.49\textwidth]{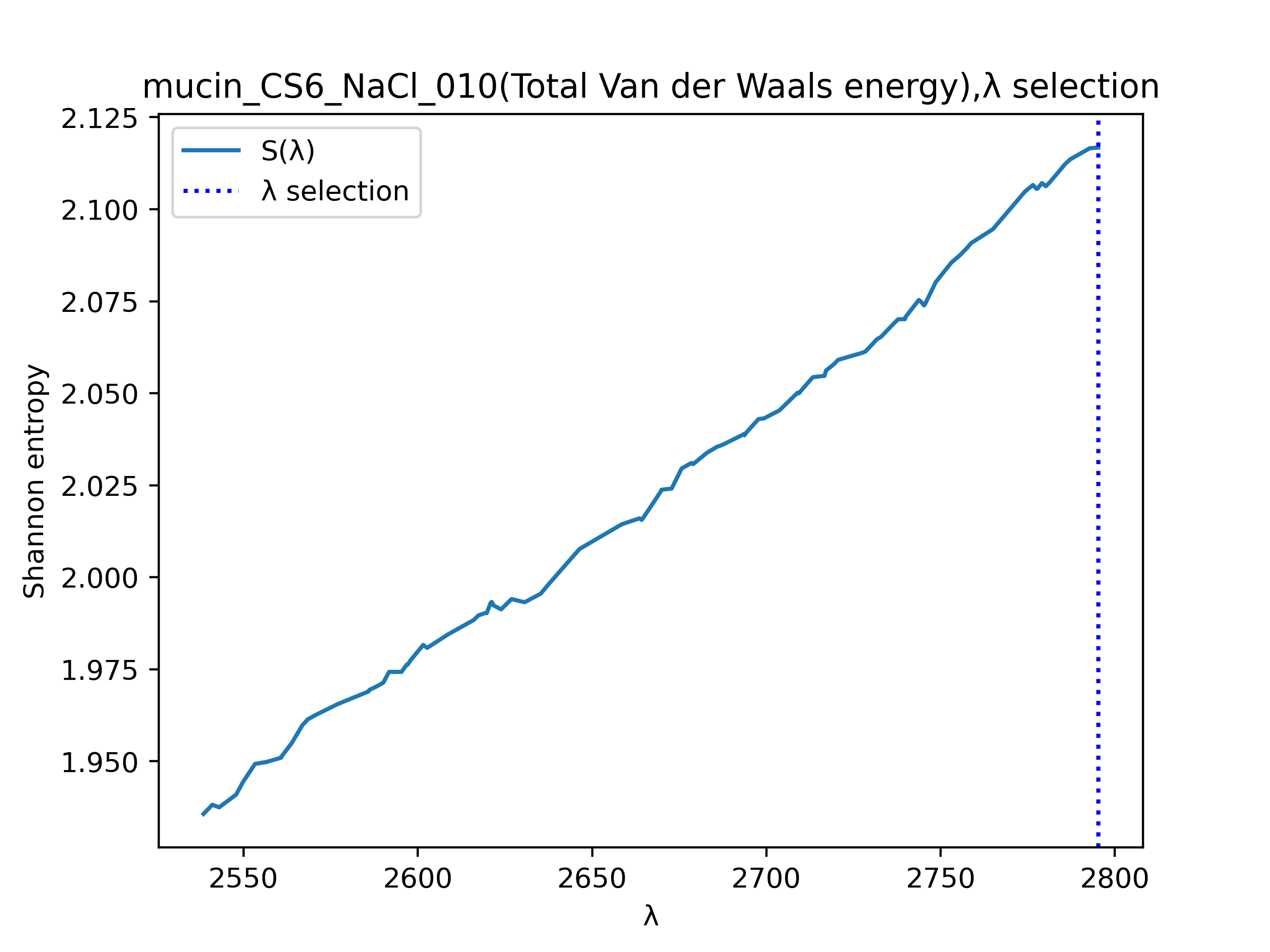}
    \caption{ Determination of $\lambda$ by the maximisation of the Shannon entropy (total Van der Waals energy of mucin with CS-6 was chosen in both cases). Observe, that in each case we got highest possible $\lambda$. Observe the order of magnitude difference in entropy between MgCl$_2$ case and NaCl case. We observe, that for all data we have entropy either close to $0.2$, what correspond with delay time $T = 2$; or close to $2.1$, what correspond with $T = 1$. }\label{fig::get_lambda}
\end{figure}

\noindent{\bf Output:}
\begin{itemize}
\item return $\lambda$   
\end{itemize}
The entropy maximization plots are presented in Fig.~\ref{fig::get_lambda}. In all cases, we got the highest possible $\lambda$ given the above-mentioned ranges of a number of ones and the recurrence rate.

\section{Results and discussion}\label{sec::results}

Entropy defined by Eq.~\eqref{eq::entropy}
reflects the complexity of the recurrence plot in respect of the diagonal lines \cite{MarwanRomanoThielKurths2007}. 
If we were to analyze the uncorrelated noise, the value of Shannon entropy would be small, which implies its low complexity. Referring, however, to Fig.~\ref{fig::get_lambda} we can see that the particular entropy value is highly dependent on the delay time; hence we can not compare entropy quantitatively for various experimental settings. It is why we focus on the recurrence plot analysis.

\begin{figure}
\centering
    \includegraphics[width=0.49\textwidth]{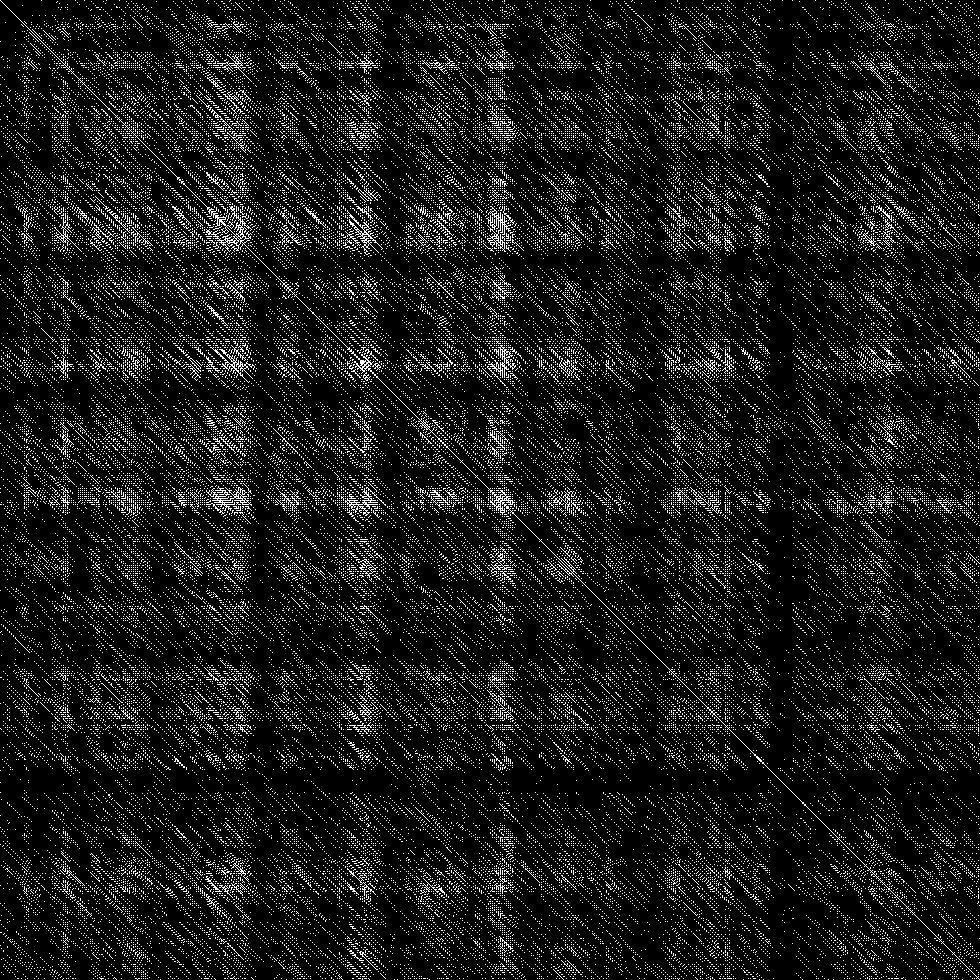}
    \includegraphics[width=0.49\textwidth]{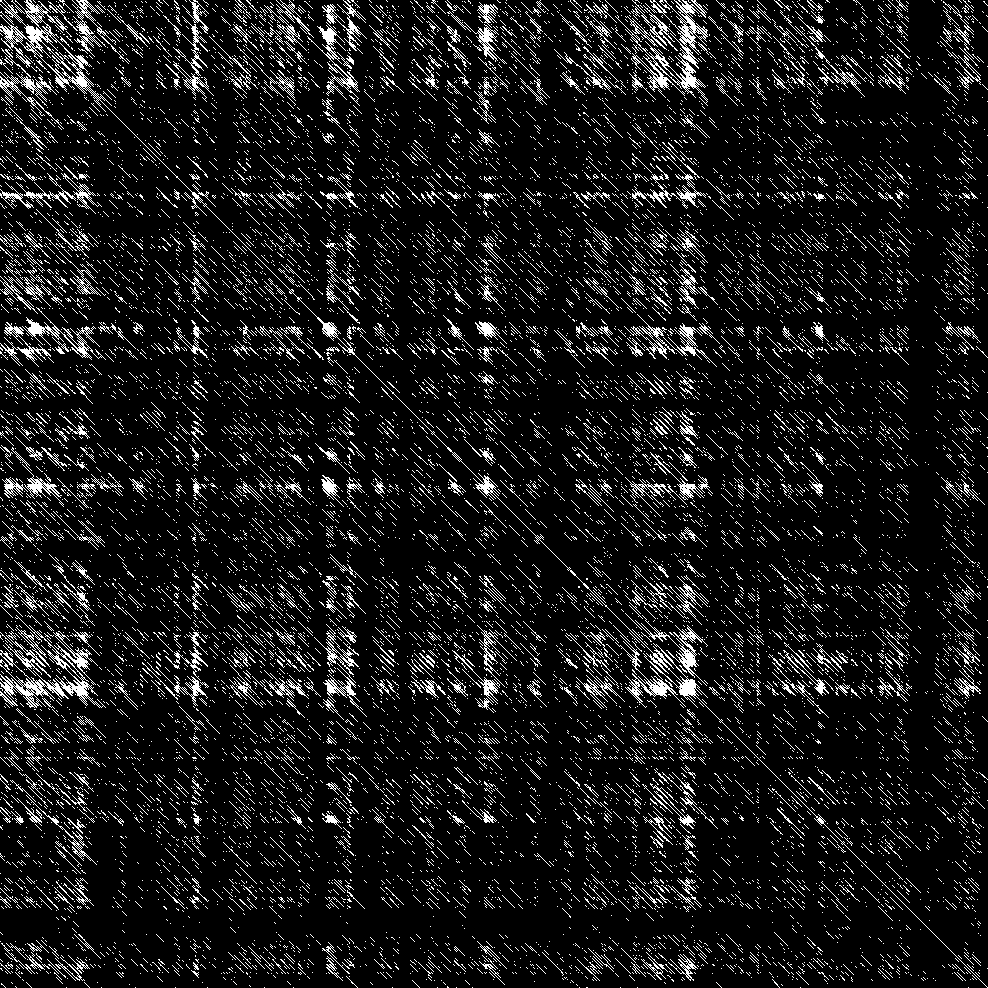}
    \includegraphics[width=0.49\textwidth]{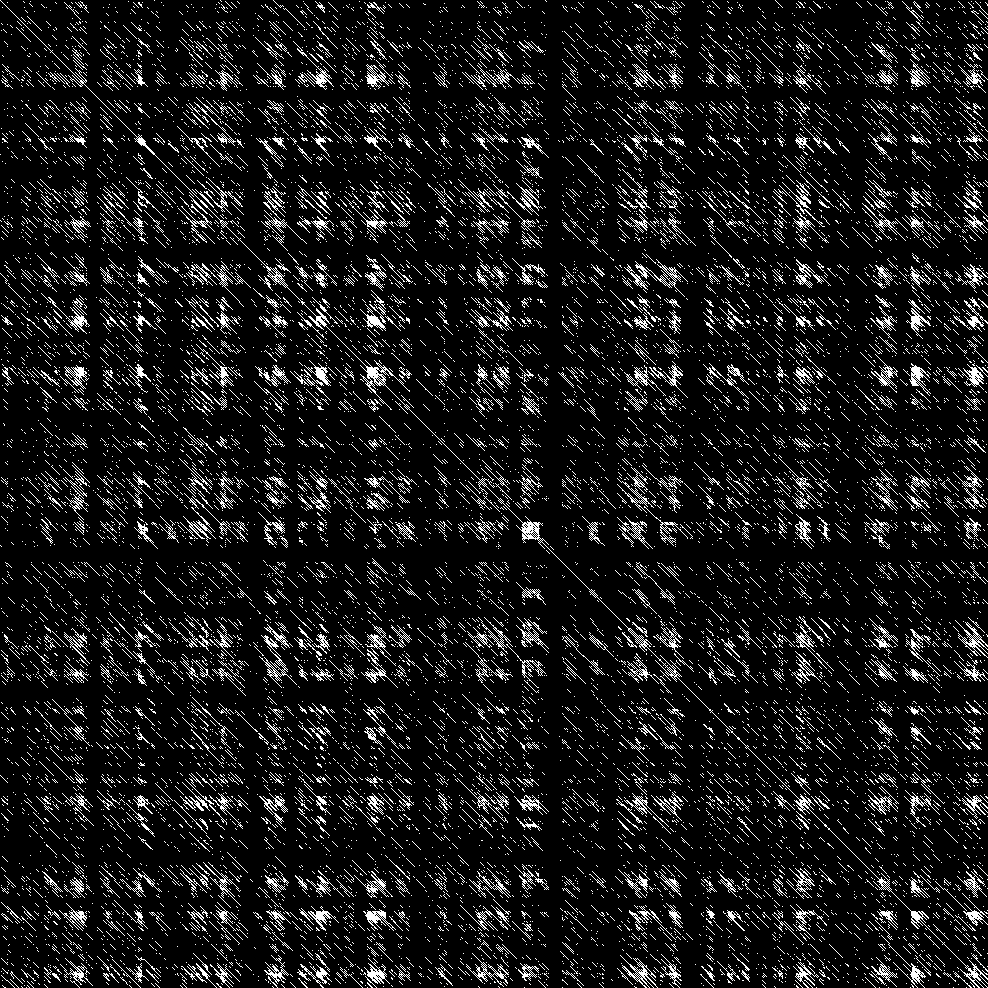}
    \includegraphics[width=0.49\textwidth]{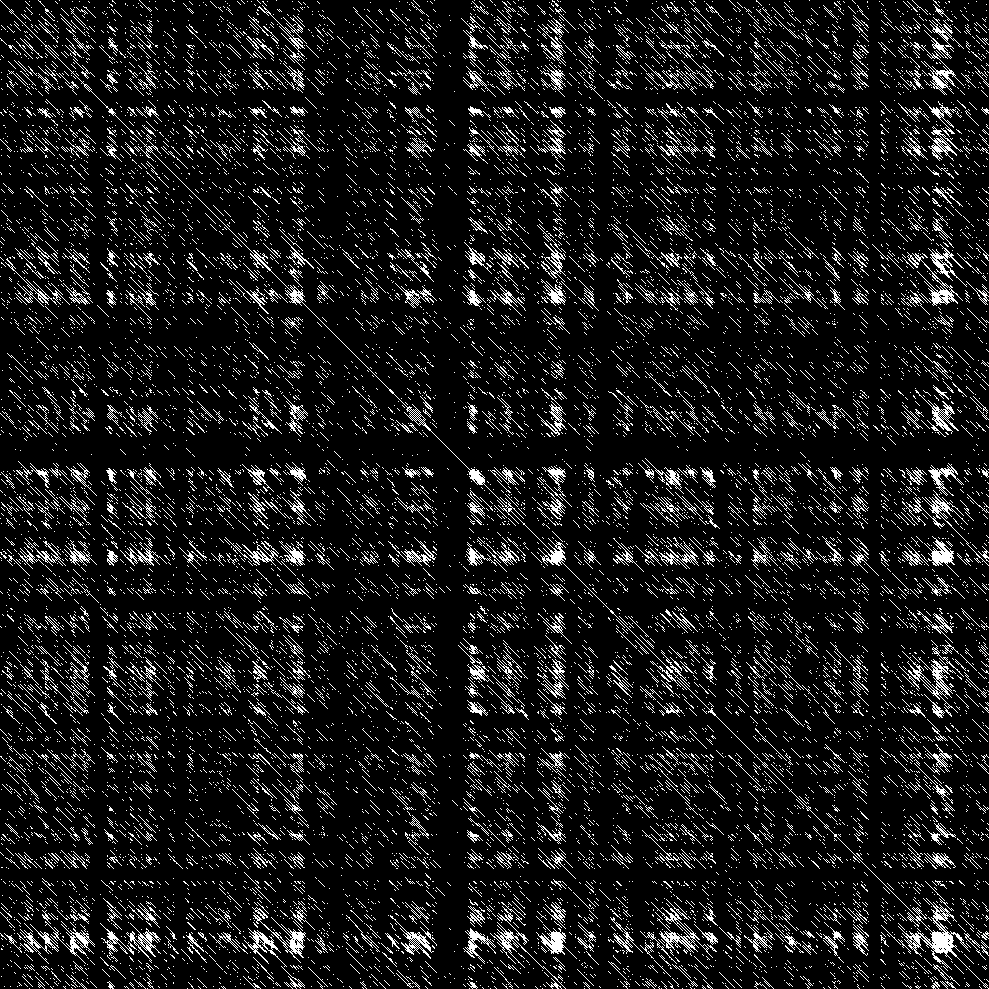}
    \caption{ Recurrence plots  mucin CS-6 with MgCl$_2$ - upper panel, CaCl$_2$ - lower panel (total Van der Waals energy feature); two different realisations of the experiment are in columns. Zeros are in white ones are black. The upper left panel (CS-6, MgCl$_2$, realisation n.o. $6$) differs (in sharpness) a bit from others, however in that particular case we have the \emph{time delay} $T = 2$, in contrary to other cases we have $T = 1$. Despite this sharpness differences, the upper panel (MgCl$_2$) is less granular (white rectangles are larger), what may suggest different dynamics.  }\label{fig::recurrence_plots_cs6}
\end{figure}

\begin{figure}
\centering
    \includegraphics[width=0.49\textwidth]{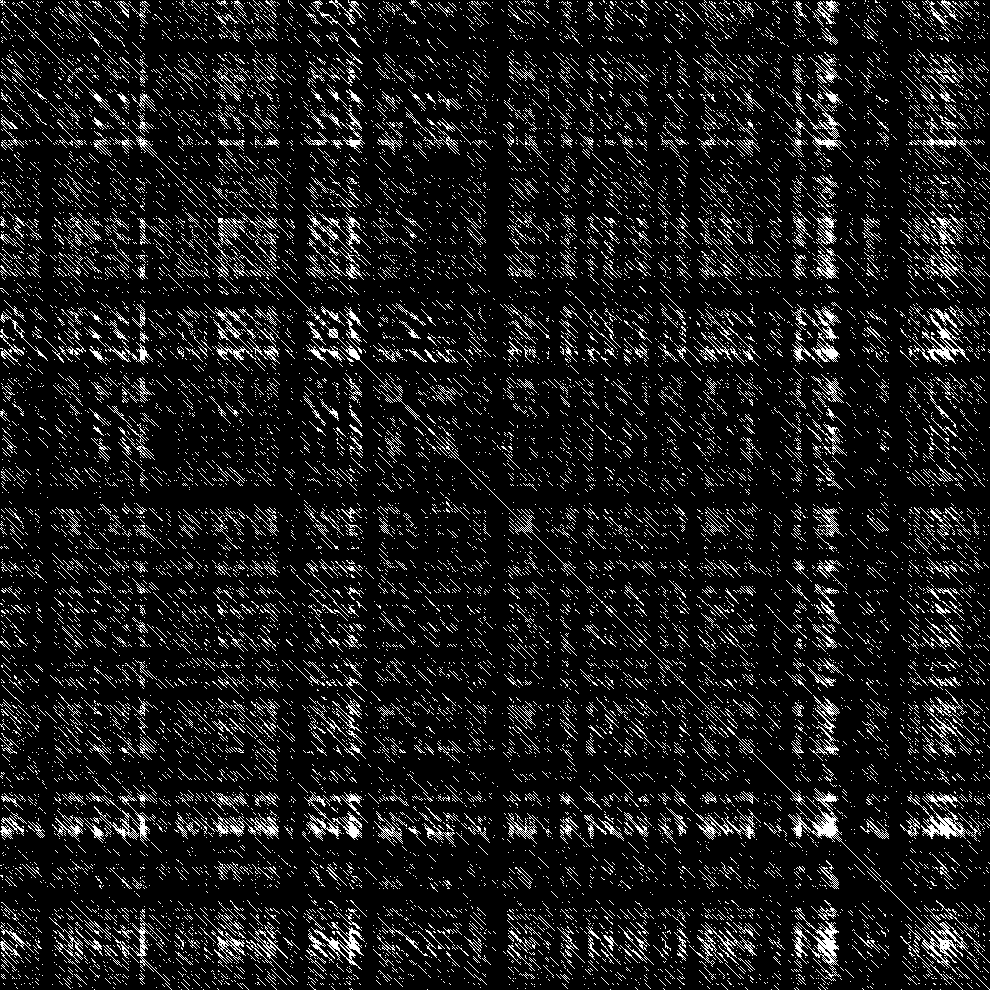}
    \includegraphics[width=0.49\textwidth]{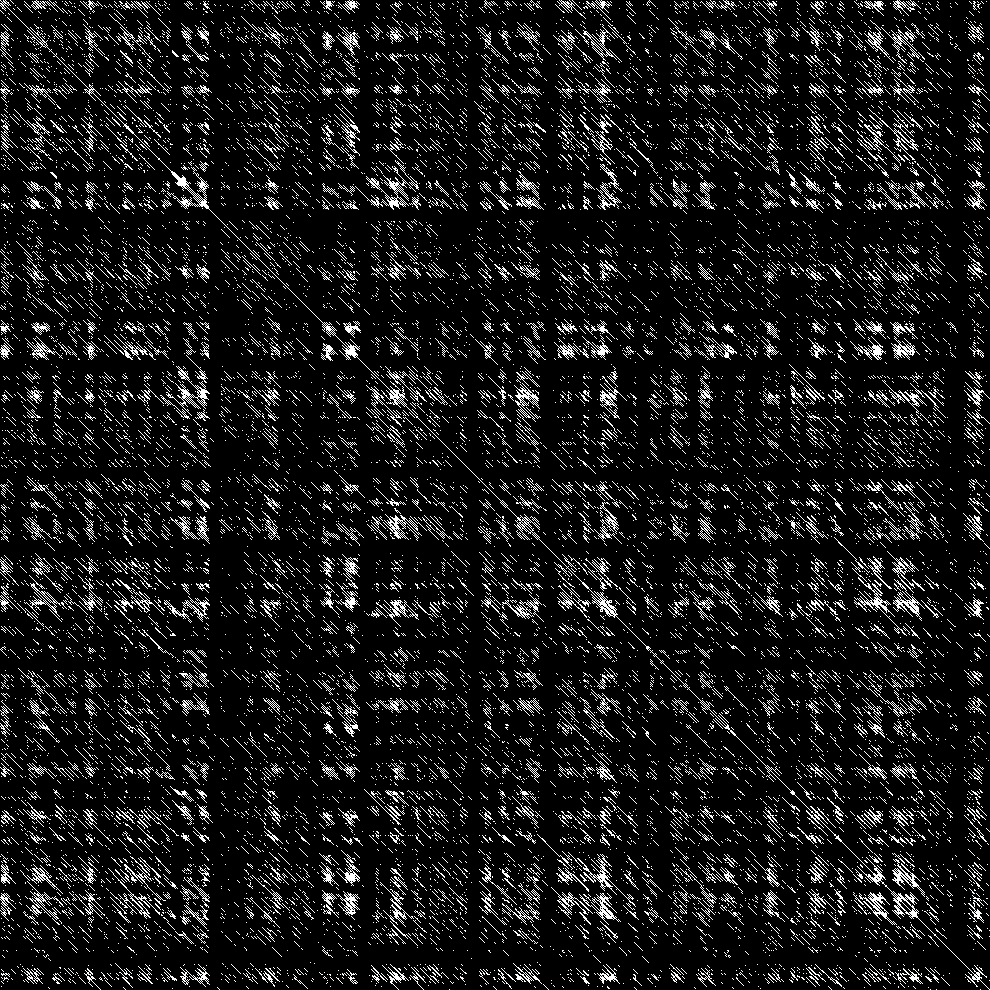}
    \includegraphics[width=0.49\textwidth]{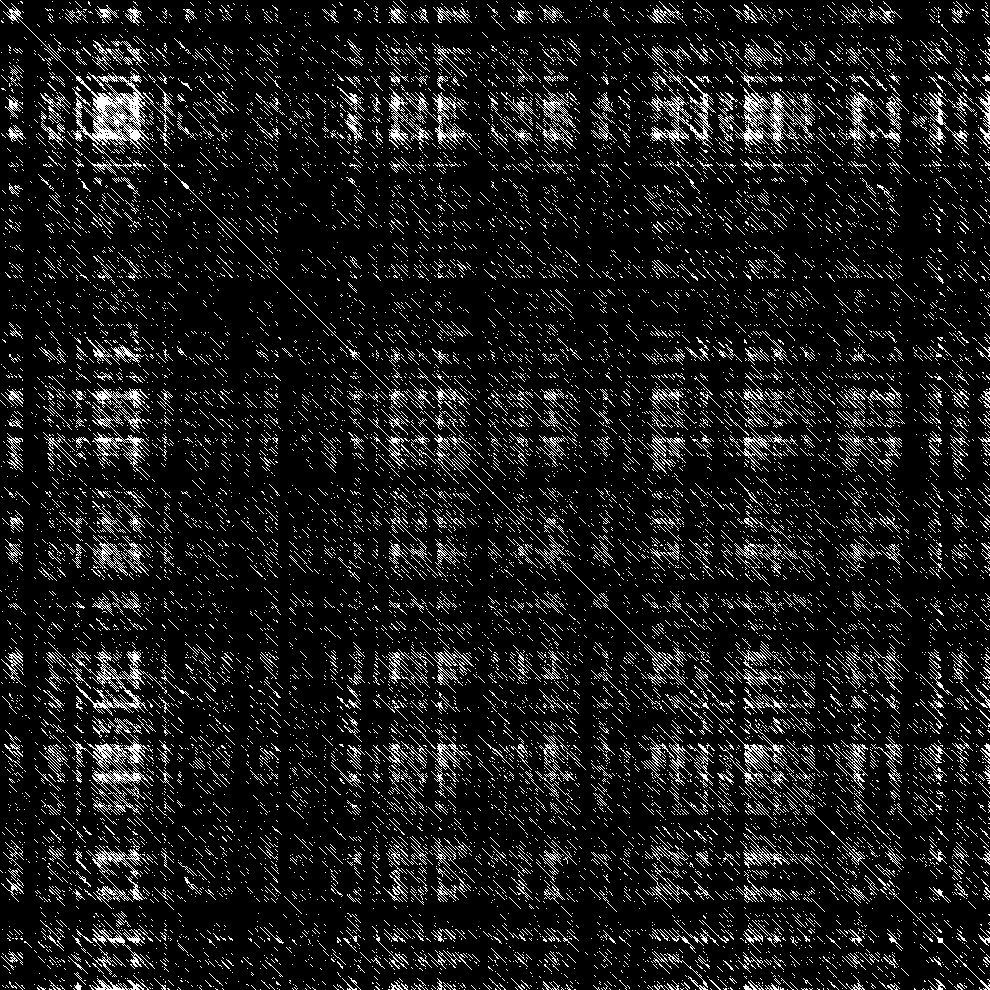}
    \includegraphics[width=0.49\textwidth]{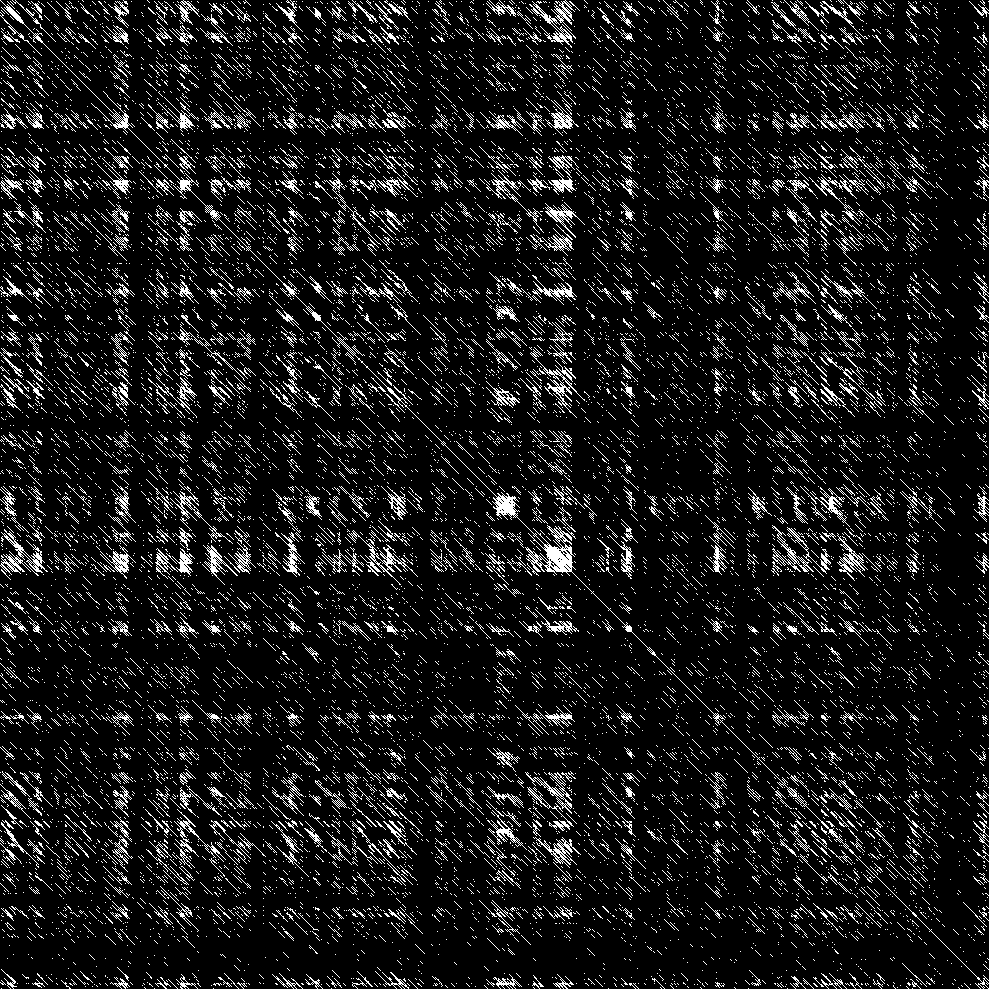}
    \caption{ Recurrence plots  mucin CS-6 with NaCl - upper panel, KCl - lower panel (total Van der Waals energy feature); two different realisations of the experiment are in columns. Zeros are in white ones are black. }\label{fig::recurrence_plots_nak_cs6}
\end{figure}

\begin{figure}
\centering
    \includegraphics[width=0.49\textwidth]{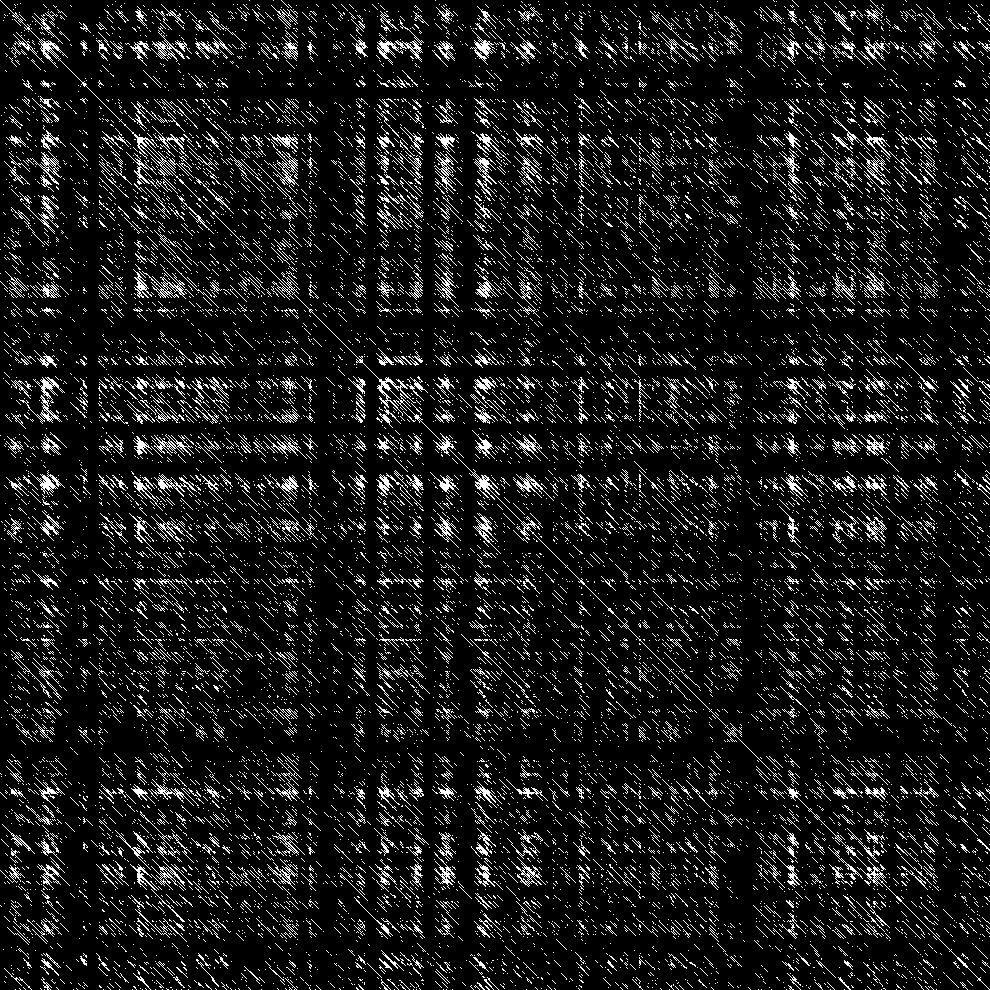}
    \includegraphics[width=0.49\textwidth]{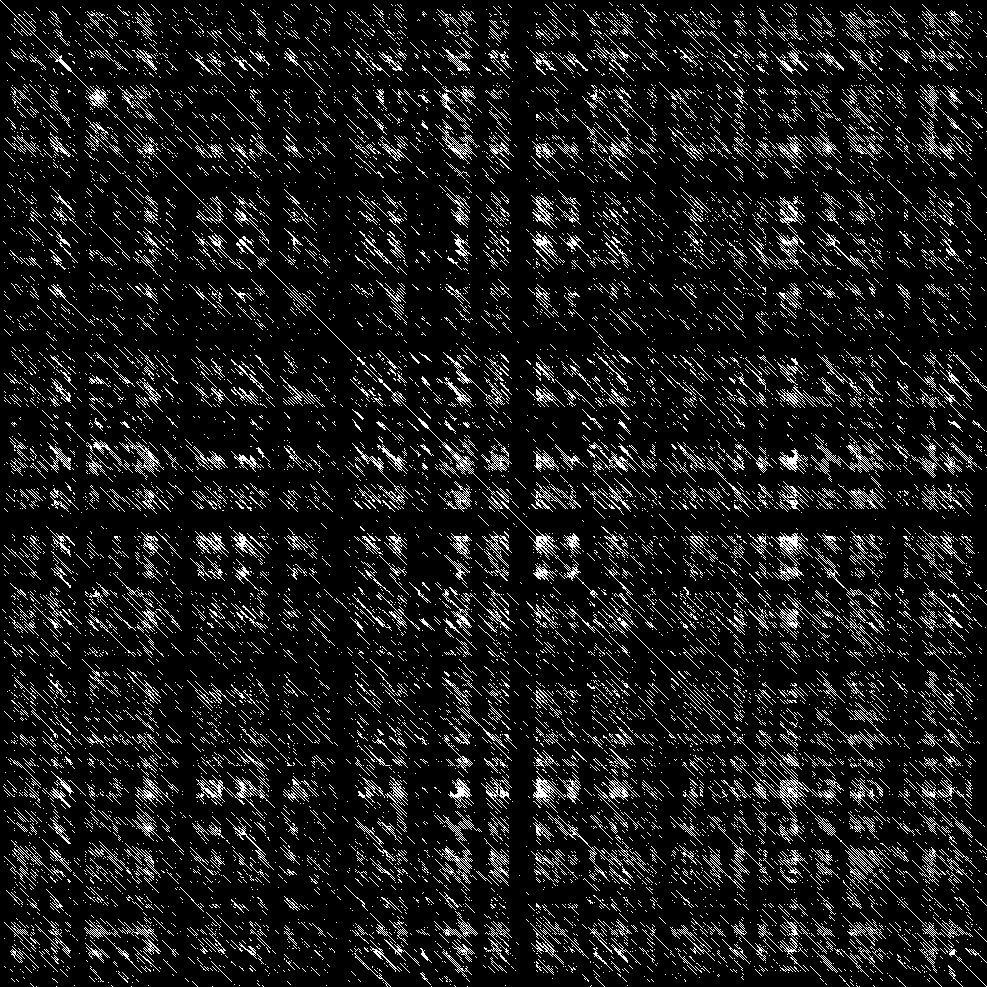}
    \includegraphics[width=0.49\textwidth]{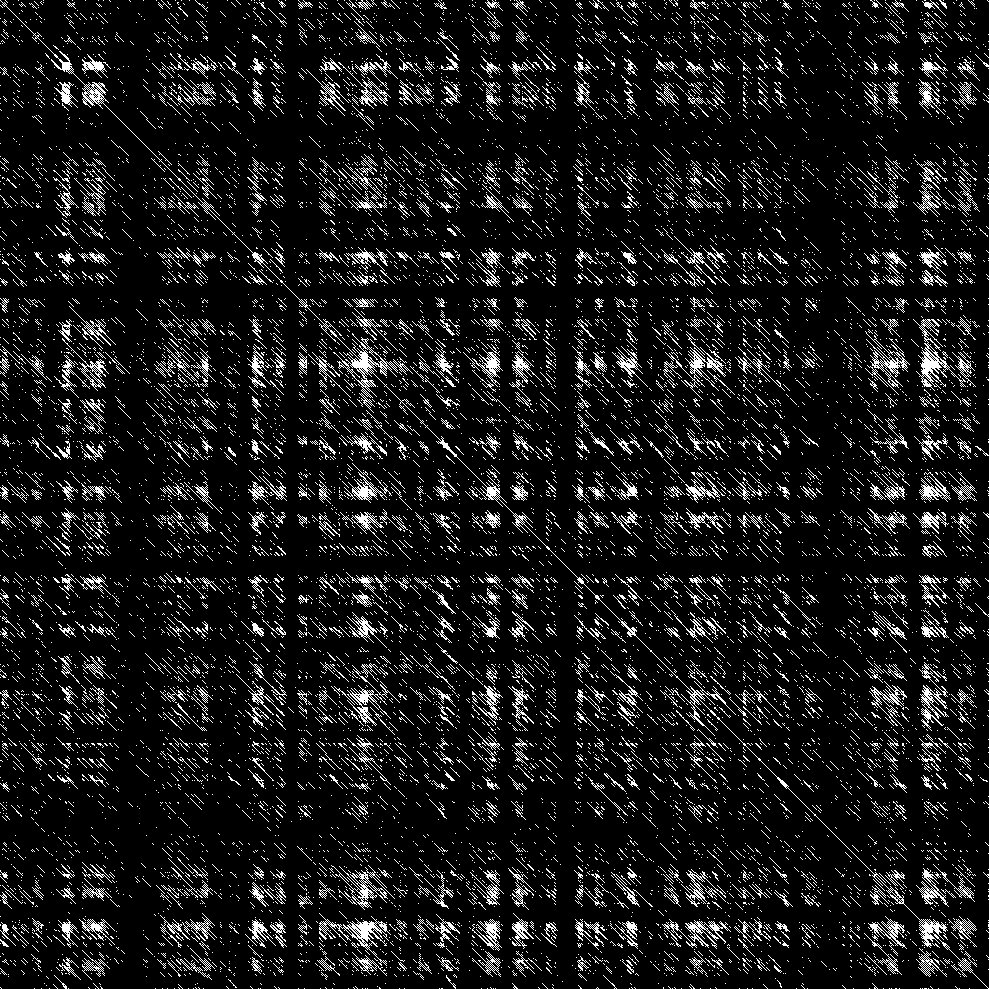}
    \includegraphics[width=0.49\textwidth]{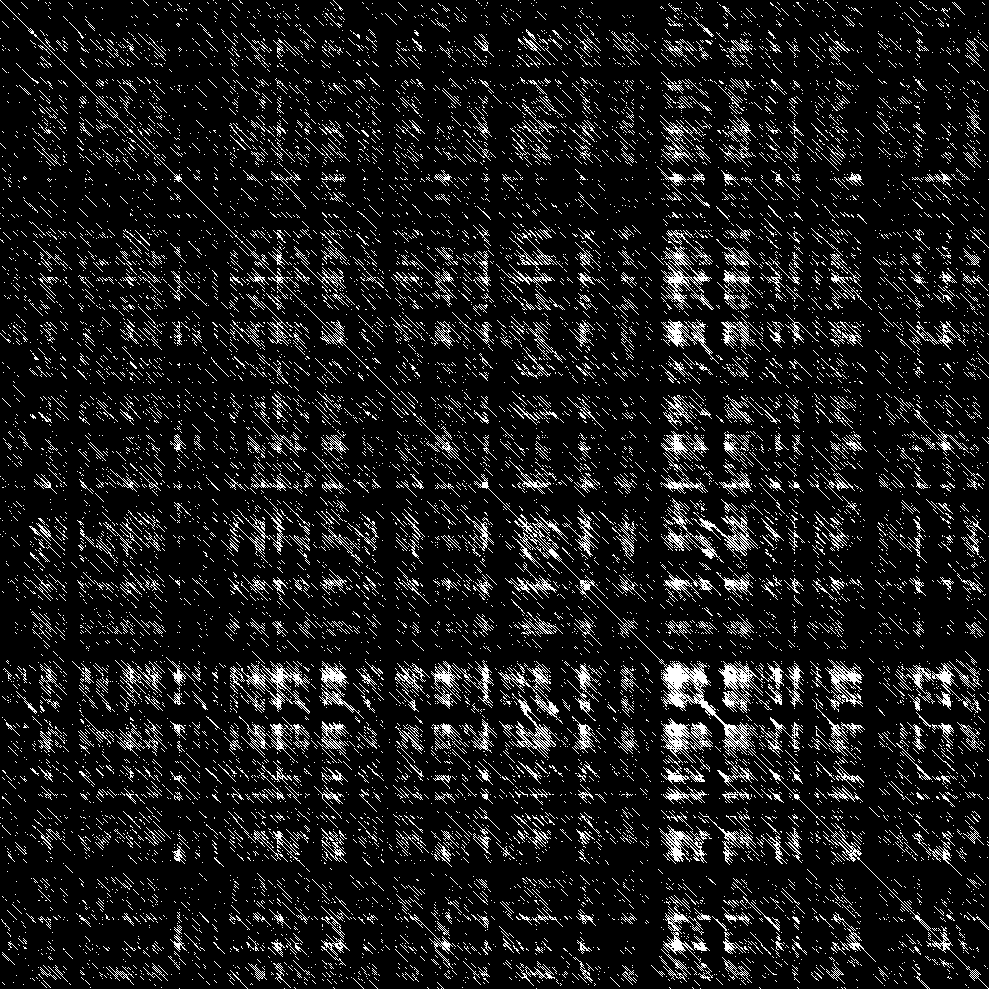}
    \caption{ Recurrence plots  mucin CS-4 with MgCl$_2$ - upper panel, CaCl$_2$ - lower panel (total Van der Waals energy feature); two different realisations of the experiment are in columns. Zeros are in white ones are black. }\label{fig::recurrence_plots_cs4}
\end{figure}

\begin{figure}
\centering
    \includegraphics[width=0.49\textwidth]{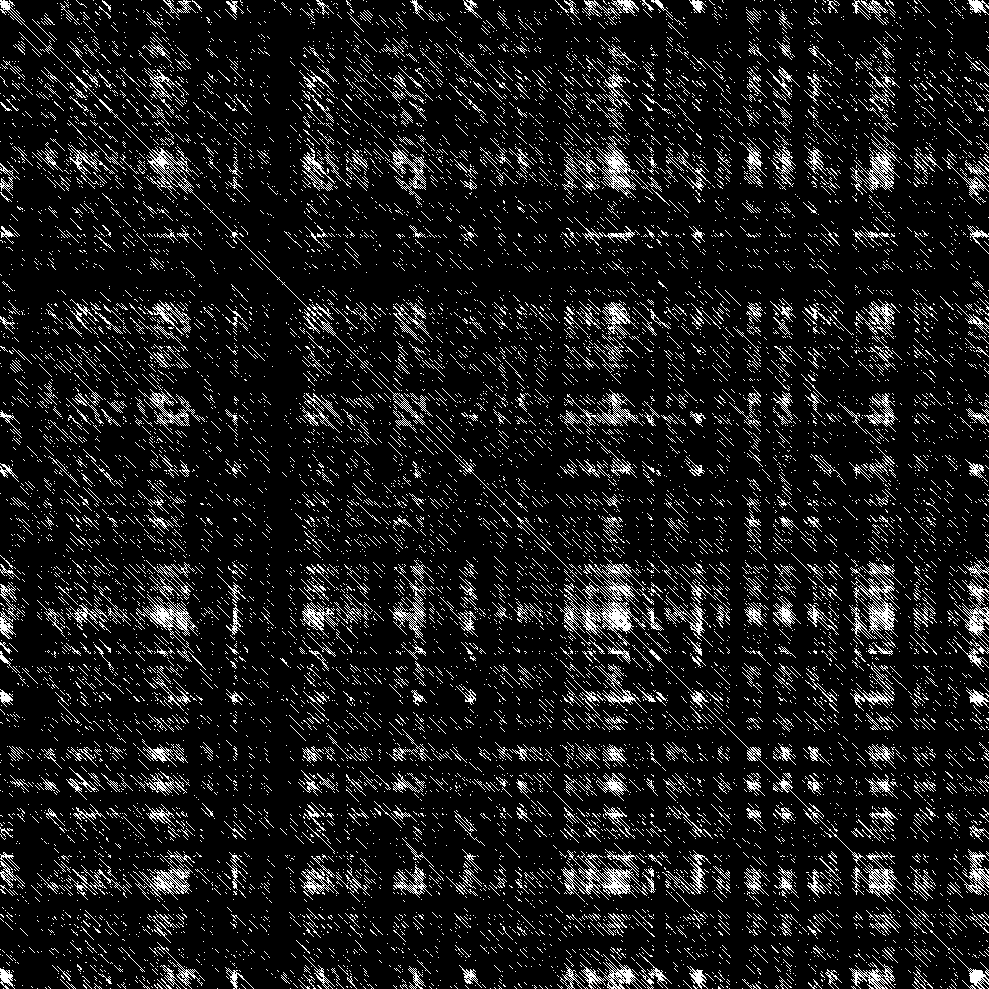}
    \includegraphics[width=0.49\textwidth]{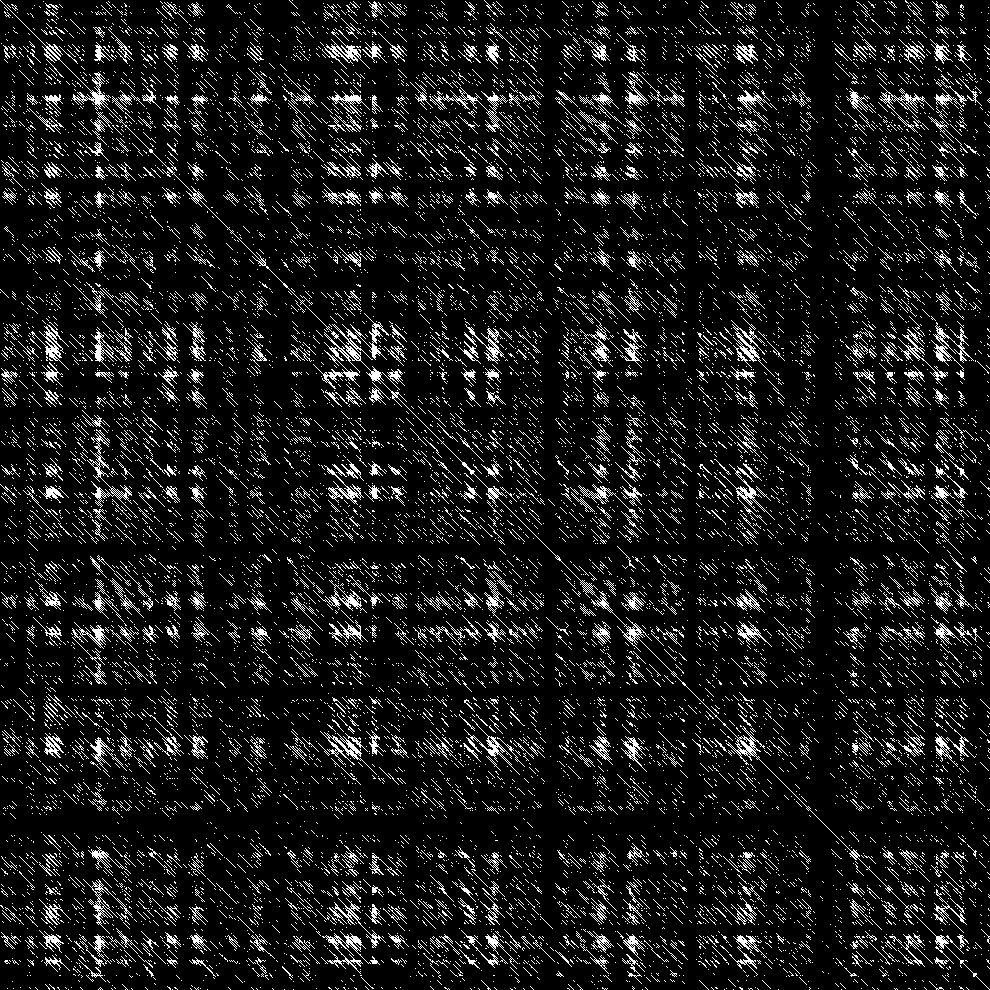}
    \includegraphics[width=0.49\textwidth]{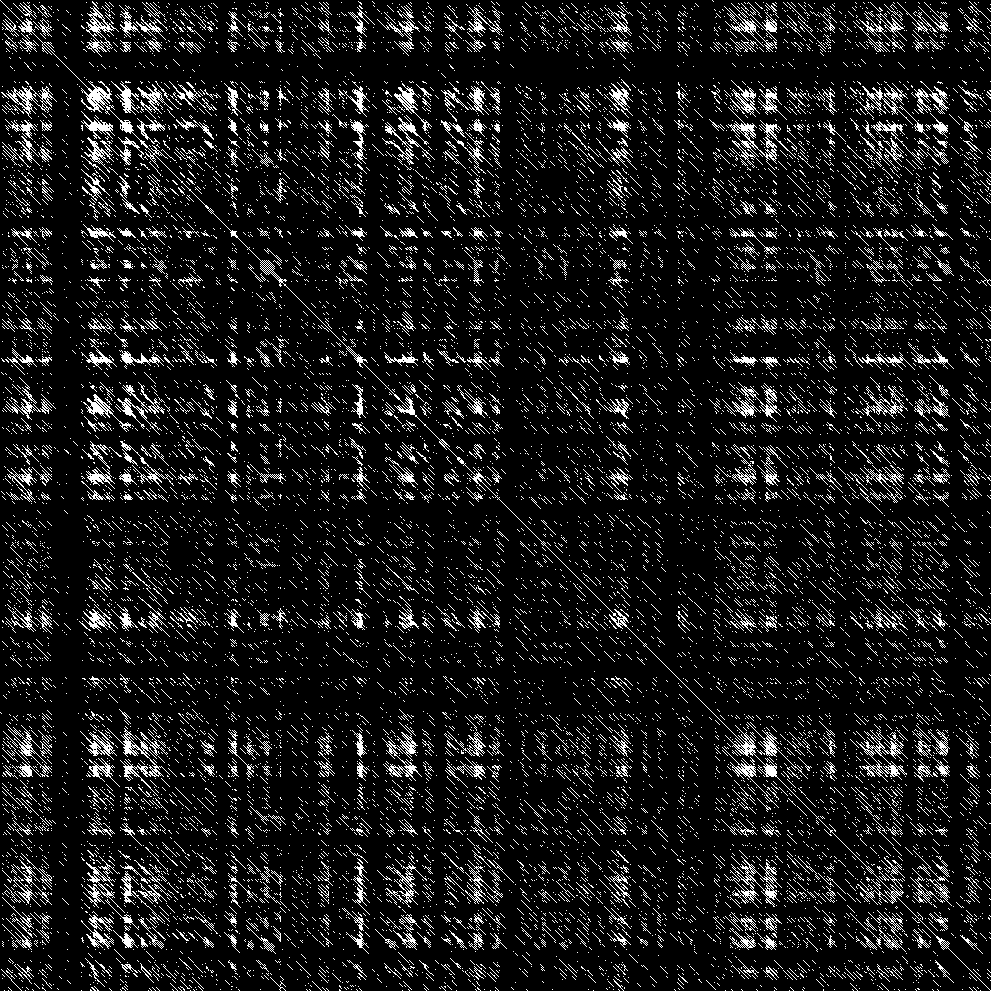}
    \includegraphics[width=0.49\textwidth]{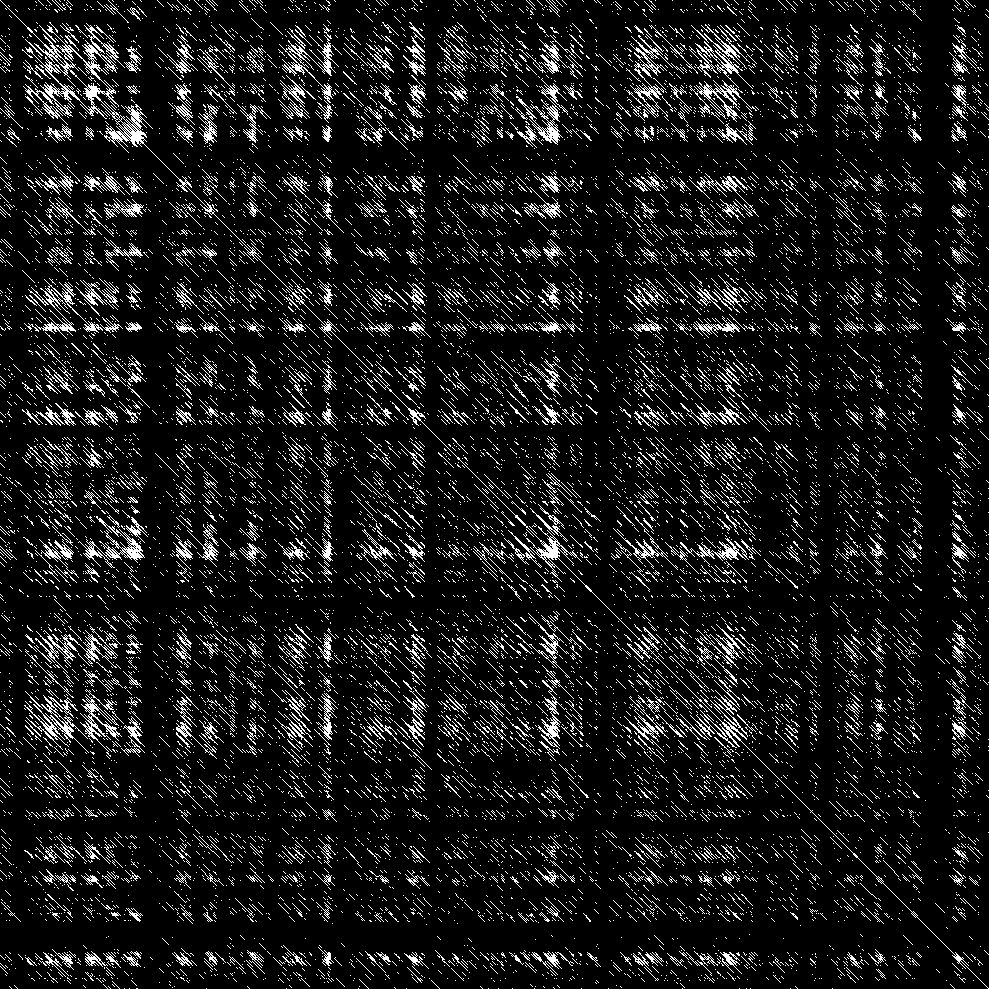}
    \caption{ Recurrence plots  mucin CS-4 with NaCl - upper panel, KCl - lower panel (total Van der Waals energy feature); two different realisations of the experiment are in columns. Zeros are in white ones are black. }\label{fig::recurrence_plots_nak_cs4}
\end{figure}

Recurrence plots for various mucies and salts (for total Van der Waals energy feature) are presented in Fig.~\ref{fig::recurrence_plots_cs6} and Fig.~\ref{fig::recurrence_plots_nak_cs6} for CS-6; and in Fig~\ref{fig::recurrence_plots_cs4}, Fig~\ref{fig::recurrence_plots_nak_cs4} for CS-4. The accurate investigation of these plots may suggest that mucin CS-6 with MgCl$_2$ are a bit different from others (there are large white rectangles on it). This observation may suggest different dynamics of this complex. Nevertheless, these plots are similar and difficult to distinguish visually. This observation was expected as we analyzed the feature (total Val der Waals energy) highly disturbed by the water-based noise. However, slight differences for one of the experimental settings may suggest that our method can extract information even for such noised data. Hereafter we point the direction for the further analysis (automatic) of recurrence plots.

The proposed method is based on the observation that one may expect some fractal patterns on the recurrence plots because of the non-linear dynamic of the system from which data and then recurrence plots are obtained. To analyze such black and white images that may have some fractal patterns, one can apply the random walk and Hurst exponent approach~\cite{zghidi2015image}.
There, one would simulate the random walker allowed in each step to move left, right, up, and down but only on black pixels (given the white pixel, the walker stays where it is). Then one would analyze the scaling (Hurst) exponent of the mean squared displacement of the walker with the number of steps. For each image, one can get the histogram of such exponents for various starting points of the random walker. Such histograms can be compared between images.
This approach has already been used successfully to analyze and classify similar black and white images of textiles fibres~\cite{ehrmann2015examination}. Let us also mention the more sophisticated multi-fractal version of the Hurst exponent approach~\cite{blachowicz2016statistical}.

\section{Conclusions}\label{sec::conclusions}
The chapter presented the recurrence plot method to perform a dynamic coarse-graining analysis of a very complex system with thousands of degrees of freedom.
The Van der Waals energy calculations show that we have reduced the number of meaningful records to just a dozen, mainly by the embedded dimension approach. Similar analysis can be performed for any other features from the numerical simulations or the real experiment. 
 
We also have noticed that the Shannon entropy computed for the recurrence plots is highly dependent on the $\tau$ (delay time) parameter. Therefore, it is crucial to choose $\tau$ gently. 

The chosen feature is highly disturbed by the water-based noise; the intended approach to testing whether the recurrence plots approach would reflect some particular characteristics of molecules despite this noise. For example, if someone looks carefully at recurrence plots, one may observe that the plot for mucin CS-6 with MgCl$_2$ looks a bit differently than others. Although recurrence plots look visually similar (due to high noise) and the Shannon entropy approach may be problematic, it is worth the effort to explore the automatic methods of further analysis. These methods should be dedicated to proceeding with black and white images of recurrence plots and are expected to reveal tiny differences and similarities in these images. The random walk and the Hurst exponent approach may be a good starting point.

\begin{acknowledgement}
This work is supported by grants of National Science Centre in Poland: (Miniatura Grant) 2019/03/X/ST3/01482 (PW), and (Sonata Bis Grant) 2016/22/E/ST6/00062 (KD). Calculations were carried out at Centre of Informatics, Tri-City Academic Supercomputer and networK (CI TASK) in Gdańsk.The work is also supported by BN-10/19 of the Institute of Mathematics and Physics of the Bydgoszcz University of Science and Technology.
\end{acknowledgement}
%
\addcontentsline{toc}{section}{Appendix}
\bibliographystyle{ieeetr}

\bibliography{references_mucin}

\end{document}